\documentclass[pra,twocolumn,letterpaper,superscriptaddress,nofootinbib,eqsecnum]{revtex4}
\usepackage{times,graphicx,graphics,amsmath,amssymb,bm}

\def\Tr{{\rm Tr}}
\newcommand \norm[1]{\|#1\|}

\newcommand \avg[1]{\left\langle #1 \right\rangle}
\newcommand{\dg}{\dagger}
\newcommand{\ket}[1]{\left|#1\right\rangle}
\newcommand{\bra}[1]{\left\langle#1\right|}

\def\lmax{\lambda_{\rm max}}
\def\lmin{\lambda_{\rm min}}
\def\Lmax{\Lambda_{\rm max}}
\def\Lmin{\Lambda_{\rm min}}
\def\abslmax{|\lambda|_{\rm max}}
\def\normH{\norm{H}}
\def\normh{\norm{h}}
\def\lambdabar{\bar\lambda}

\def\cO{{\cal O}}
\def\({\left(}
\def\){\right)}
\def\vec#1{\bm{#1}}

\begin{document}

\title{Quantum-limited metrology with product states}

\author{Sergio Boixo}
\affiliation{Department of Physics and Astronomy, MSC07-4220,
University of New Mexico, Albuquerque, New Mexico 87131-0001}
\affiliation{Los Alamos National Laboratory, Los Alamos, New Mexico 87545, USA}

\author{Animesh Datta}
\affiliation{Department of Physics and Astronomy, MSC07-4220,
University of New Mexico, Albuquerque, New Mexico 87131-0001}

\author{Steven T. Flammia}
\email{sflammia@perimeterinstitute.ca}
\affiliation{Department of Physics and Astronomy, MSC07-4220,
University of New Mexico, Albuquerque, New Mexico 87131-0001}
\affiliation{Perimeter Institute for Theoretical Physics, Waterloo, Ontario N2L 2Y5,
Canada}

\author{Anil Shaji}
\affiliation{Department of Physics and Astronomy, MSC07-4220,
University of New Mexico, Albuquerque, New Mexico 87131-0001}

\author{Emilio Bagan}
\affiliation{Grup de F\'{i}sica Te\`{o}rica, Universitat Aut\`{o}noma de
Barcelona, 08193 Bellaterra (Barcelona), Spain}
\affiliation{Department of Physics and Astronomy, MSC07-4220,
University of New Mexico, Albuquerque, New Mexico 87131-0001}

\author{Carlton M. Caves}
\affiliation{Department of Physics and Astronomy, MSC07-4220,
University of New Mexico, Albuquerque, New Mexico 87131-0001}
\affiliation{Department of Physics, University of Queensland,
Brisbane, Queensland 4072, Australia}

\date{\today}

\begin{abstract}
  We study the performance of initial product states of $n$-body
  systems in generalized quantum metrology protocols that involve
  estimating an unknown coupling constant in a nonlinear $k$-body
  ($k\ll n$) Hamiltonian.  We obtain the theoretical lower bound on
  the uncertainty in the estimate of the parameter.  For arbitrary
  initial states, the lower bound scales as $1/n^k$, and for initial
  product states, it scales as $1/n^{k-1/2}$.  We show that the latter
  scaling can be achieved using simple, separable measurements.  We
  analyze in detail the case of a quadratic Hamiltonian ($k=2$),
  implementable with Bose-Einstein condensates.  We formulate a simple
  model, based on the evolution of angular-momentum coherent states,
  which explains the $\cO(n^{-3/2})$ scaling for $k=2$; the model
  shows that the entanglement generated by the quadratic Hamiltonian
  does not play a role in the enhanced sensitivity scaling.  We show
  that phase decoherence does not affect the $\cO(n^{-3/2})$
  sensitivity scaling for initial product states.
\end{abstract}
\pacs{03.65.Ta,03.67.-a,03.67.Lx ,06.20.Dk}

\maketitle

\section{Introduction}

Parameter estimation is a fundamental physical task.  It typically
involves picking a physical system whose state, through its
evolution, depends on the value of the parameter.  In most quantum
metrology
schemes~\cite{wineland94a,bollinger96a,huelga97a,childs00a,dunningham04a,
luis04a,burgh05a,cappellaro05a,beltran05a,giovannetti06a,roy06a,boixo07a,
knill07a,luis07a,rey07a,boixo07b,partner07a,choi07a}, this system,
which we call the ``probe,'' is a composite made up of $n$ elementary
quantum constituents.  The influence of the unknown parameter
$\gamma$ on the probe is described by an $n$-body Hamiltonian
\begin{equation}
H_\gamma=\gamma H\;,
\end{equation}
in which $\gamma$ appears as a coupling constant and $H$ is a
dimensionless coupling Hamiltonian (we use units with $\hbar=1$, so
$\gamma$ has units of frequency).  The precision with which $\gamma$
can be determined depends on the initial state of the probe, the
nature of the parameter-dependent Hamiltonian, and the measurements
that are performed on the probe to extract information about the
parameter.  Other factors, such as decoherence in the
probe~\cite{huelga97a,boixo06a,gilbert06a,shaji07a}, also have an
effect on the achievable sensitivity.

The appropriate measure of the precision with which $\gamma$ can be
determined is the units-corrected mean-square deviation of the
estimate $\gamma_{\rm est}$ from the true value
$\gamma$~\cite{braunstein94a,braunstein96a}:
\begin{equation}
\delta\gamma=
\biggl\langle\biggl(
\frac{\gamma_{\rm est}}{|d\langle\gamma_{\rm est}\rangle/d\gamma|}
-\gamma
\biggr)^{\!2}
\biggr\rangle^{1/2}\;.
\label{eq:deltagamma}
\end{equation}
This estimator uncertainty is inversely proportional to the
displacement in Hilbert space of the state of the probe corresponding
to small changes in $\gamma$.  The fundamental limit on the precision
of parameter estimation,
\begin{equation}
\label{eq:cramer-rao}
\delta \gamma \ge \frac{1}{\sqrt\nu}\frac{1}{2t\Delta H}\;,
\end{equation}
called the {\em quantum Cram\'er-Rao bound\/}
(QCRB)~\cite{helstrom76a,holevo82a,braunstein94a,braunstein96a}, is
an expression of the maximum amount the state can change under the
evolution due to $H_\gamma$.  In Eq.~\eqref{eq:cramer-rao}, $\nu$ is
the number of trials with independent, identical probes, $t$ is the
time for which each probe evolves under $H_\gamma$, and $\Delta
H=(\langle H^2\rangle-\langle H\rangle^2)^{1/2}$ denotes the
uncertainty in $H$ for each probe, which does not change under the
evolution due to $H_\gamma$.  The QCRB is independent of the choice
of estimator and is achievable asymptotically in the limit of a large
number of trials, provided the initial state of the probe is a pure
state. If the initial state of the probe is not pure or if nonunitary
processes destroy the purity of the initial state, the
bound~\eqref{eq:cramer-rao} is not tight, and a stricter version of
the QCRB, given in~\cite{helstrom76a,holevo82a,braunstein94a,
braunstein96a}, can be used, but we have no need for this stricter
bound in this paper.

The uncertainty in $H$ is bounded above by
\begin{equation}
\Delta H\le\frac{\Lmax-\Lmin}{2}\;,
\end{equation}
where $\Lmax$ and $\Lmin$ are the maximum and minimum eigenvalues of
$H$.  The difference between the largest and least eigenvalues,
denoted by
\begin{equation}
\normH=\Lmax-\Lmin\;,
\end{equation}
is an operator semi-norm of $H$.  Using this semi-norm, we can
write a state-independent version of the QCRB~\cite{braunstein94a}:
\begin{equation}
\label{eq:semi-norm}
\delta \gamma \ge \frac{1}{\sqrt\nu}\frac{1}{t\normH} \; .
\end{equation}

This bound can be achieved by using the initial state
$(\ket{\Lmax}+\ket{\Lmin})/\sqrt 2$, which evolves after time~$t$ to
\begin{align}
e^{-iH_\gamma t}\frac{1}{\sqrt2}&(\ket{\Lmax}+\ket{\Lmin})\nonumber\\
&=\frac{1}{\sqrt2}(e^{-i\Lmax\gamma t}\ket{\Lmax}+e^{-i\Lmin\gamma t}\ket{\Lmin})\;.
\end{align}
Measurement in a basis that includes the states
$\ket{\pm}=(\ket{\Lmax}\pm\ket{\Lmin})/\sqrt2$ yields outcomes $\pm1$
with probabilities $p_+=\cos^2(\normH\gamma t/2)$ and
$p_-=\sin^2(\normH\gamma t/2)$.  Letting $\sigma$ denote an
observable with the outcome values $\pm1$, we have
$\avg{\sigma}=\cos(\normH\gamma t)$ and
$\Delta\sigma=|\!\sin(\normH\gamma t)|$.  An appropriate estimator is
defined in terms of the mean of the outcomes,
\begin{align}
\frac{1}{\nu}\sum_{k=1}^\nu\sigma_k&\equiv
\cos(\normH\gamma_{\rm est}t)\nonumber\\
&=\cos(\normH\gamma t)-\normH\,t\sin(\normH\gamma t)(\gamma_{\rm est}-\gamma)\;,\nonumber\\
\label{eq:gammaest}
\end{align}
where the second expression is the linear approximation to the
relation between $\gamma_{\rm est}$ and the mean of the outcomes,
holding statistically in the limit of a large number of trials,
specifically, $\nu\gg\tan^2(\normH\gamma t)$.  Now it is easy to see
that $\avg{\gamma_{\rm est}}=\gamma$ and
\begin{align}
\delta\gamma&=\avg{(\gamma_{\rm est}-\gamma)^2}^{1/2}\nonumber\\
&=\frac{\Delta\sigma}{\sqrt\nu}\frac{1}{\normH\,t\,|\!\sin(\normH\gamma t)|}\nonumber\\
&=\frac{1}{\sqrt\nu}\frac{1}{t\normH}\;,
\end{align}
showing that the bound~\eqref{eq:semi-norm} can be achieved and thus
making it the fundamental limit to quantum metrology.

The $\sqrt\nu$ factor in Eqs.~\eqref{eq:cramer-rao} and
\eqref{eq:semi-norm} is the well-understood statistical improvement
available from averaging over many probes.  In the remainder of the
paper, we sometimes do not include this factor explicitly, referring
to the remaining term on right-hand side of Eq.~\eqref{eq:cramer-rao}
or Eq.~\eqref{eq:semi-norm} as the QCRB, always remembering, of
course, that generally the bound can only be achieved asymptotically
in the limit of large $\nu$.

It is clear from Eqs.~\eqref{eq:cramer-rao} and~\eqref{eq:semi-norm}
that strategies for improving the precision in estimating a parameter
include changing the initial state of the probe, the coupling
Hamiltonian $H$, or both.  The case that has received most attention
in the past is the one in which the probe constituents are coupled
independently to the parameter:
\begin{equation}
\label{eq:single}
H = \sum_{j=1}^n h_j \;.
\end{equation}
Here $h_j$ is a single-body operator acting on the $j$th constituent
of the probe (hence, all these operators commute).  In this case, the
QCRB~\eqref{eq:semi-norm} scales like $\delta \gamma =  \cO(1/n)$, a
scaling known as the {\em Heisenberg
limit\/}~\cite{wineland94a,giovannetti06a}.  This scaling outperforms
that attainable with classical statistics, which goes as $\delta
\gamma = \cO( 1/\sqrt n)$, a scaling known as the {\em standard
quantum limit\/} or, sometimes, as the {\em shot-noise limit}.  The
$1/\sqrt n$ scaling is the optimum sensitivity allowed by the
QCRB~\eqref{eq:cramer-rao} when the coupling Hamiltonian has the
form~\eqref{eq:single} and the optimization of the probe state is
restricted to product states.

Achieving the quantum-enhanced Heisenberg sensitivity with the linear
coupling Hamiltonian~\eqref{eq:single} requires the probe to be
initialized in a highly entangled state, which is a formidable
challenge using current technology~\cite{wineland94a}. Much progress
has been made in preparing such states toward precisely this end,
e.g., by using measurement-induced squeezing~\cite{geremia03a}, but
it is still currently infeasible to use Heisenberg-limited
experiments to outperform the best measurements operating at the
standard quantum limit, although Heisenberg-like scalings have been
achieved in related serial protocols that involve repeated
interactions with a single constituent and thus do not involve
entanglement~\cite{higgins07a}.  Practical
proposals~\cite{toscano06a} for reaching the Heisenberg limit have
also been made in the context of measurements on a harmonic
oscillator prepared in a state that displays sub-Planck phase-space
structure~\cite{zurek01a}.

More general families of Hamiltonians, containing nonlinear couplings
of the constituents to the parameter, in contrast to the independent,
linear coupling of Eq.~\eqref{eq:single}, can perform better than the
$1/n$ scaling of the Heisenberg limit, while respecting the
QCRB~\cite{luis04a,beltran05a,roy06a,boixo07a,luis07a,rey07a,choi07a}.
In particular, when the coupling Hamiltonian has symmetric $k$-body
terms in it, it is possible to achieve the scaling $\delta\gamma =
\cO(n^{-k})$, as shown in~\cite{boixo07a}.  This $\cO(n^{-k})$
scaling requires entangled input states, but we will show here that
the optimal scaling with initial product states is $\cO(n^{-k+1/2})$.
Thus scalings better than the $1/n$ Heisenberg scaling are possible
for $k\ge2$, even with initial product states for the probe, a result
found for $k=2$ in~\cite{luis04a,beltran05a,choi07a}.

In this paper we investigate the theoretical and practical bounds on
precision in the generalized quantum metrology scheme introduced
in~\cite{boixo07a}, which allows for nonlinear couplings of the probe
constituents to the parameter.  In particular, we study the nonlinear
coupling Hamiltonian~\cite{boixo07a}
\begin{equation}
\label{eq:hs}
  H = \bigg( \sum_{j=1}^n h_j\bigg)^k
  = \sum^n_{a_1, \ldots , a_k} h_{a_1}\cdots h_{a_k} \;.
\end{equation}
For simplicity, we assume that the probe constituents are identical
and that the single-body operators $h_j$ are the same for all the
constituents.  We review in Sec.~\ref{sec2} the optimal precision
that can be achieved when the probe can be prepared in any initial
state, particularly, entangled states of the probe constituents, but
our emphasis in this paper is on the precision that can be attained
when the initial state is restricted to be a product state.

In Sec.~\ref{sec3} we show that the optimal precision with
product-state inputs scales as $\cO(n^{-k+1/2})$, and we find the
corresponding optimal input state for the probe constituents.  The
sensitivity that can be achieved in practice depends, as mentioned
above, on the measurements that are performed on the probe to extract
information about $\gamma$.  By analyzing the short-time limit, we
show in Sec.~\ref{sec4} that when the probe is initialized in a
product state, simple, separable measurements on the probe
constituents can achieve the optimal sensitivity.  The conclusion,
reached in~\cite{luis04a,beltran05a,choi07a} for specific cases, is
that scalings better than the standard quantum limit---indeed, better
than the Heisenberg limit---can be had without the need to invest in
the generation of fragile initial entangled states.  Our scheme thus
circumvents a major bottleneck in attaining, in practice, scalings
superior to $1/\sqrt n$.

In Sec.~\ref{sec5}, we analyze in detail a quadratic coupling
Hamiltonian ($k=2$) for effective qubits, with $h_j=Z_j/2$, where
$Z_j$ is the Pauli $z$ operator for the $j$th qubit.  This case can
be implemented in Bose-Einstein condensates~\cite{rey07a,choi07a}, as
was suggested in~\cite{boixo07a}, and in fermionic atoms in optical
lattices~\cite{rey07a}.  We show that the optimal sensitivity for
input product states, scaling as $\cO(n^{-3/2})$, can be achieved by
using the optimal product-state input found in Sec.~\ref{sec3} and
making separable measurements of equatorial components of the total
angular momentum of the effective qubits. Moreover, we show that, as
was found independently by~\cite{choi07a}, the $\cO(n^{-3/2})$
scaling is attainable with these measurements starting with almost
any state of the constituents except the equatorial states, which
were the subject of the analysis in~\cite{rey07a}.  We formulate a
simple model, based on the evolution of angular-momentum coherent
states, that explains the origin of the $\cO(n^{-3/2})$ scaling.  The
model indicates that the entanglement generated by the quadratic
Hamiltonian does not play a role in the enhanced sensitivity, and it
suggests that, unlike protocols based on the use of entangled inputs,
the product-state scheme should not be extremely sensitive to
decoherence.  We verify this suggestion by a brief analysis of the
effect of phase decoherence in Sec.~\ref{sec:dec}.

Section~\ref{sec6} concludes with a brief summary of our results,
including an extension of the coherent-state model to arbitrary $k$,
and a discussion of our perception of the field of quantum-enhanced
metrology.

\section{Generalized quantum metrology and the quantum Cram\'er-Rao bound}
\label{sec2}

Attaining the QCRB~\eqref{eq:semi-norm} requires using an appropriate
initial state and making appropriate measurements to extract the
information about $\gamma$, but changing the optimal scaling with $n$
requires changing the dependence of the coupling Hamiltonian $H$ on
$n$.  This can be done by replacing the linear coupling
Hamiltonian~\eqref{eq:single}, which has just $n$ terms, with the
nonlinear Hamiltonian~\eqref{eq:hs}.  The coupling
Hamiltonian~\eqref{eq:hs} describes a system with symmetric $k$-body
couplings, including self-interactions, and it has $n^k$ terms.  For
example, if the constituents are spin-$\frac{1}{2}$ particles and the
operator $h_j$ is the $z$-component of the the $j$th particle's spin,
then $H$ describes a coupling of the parameter to the $k$th power of
the $z$-component of the total angular momentum.

The eigenvectors of $H$ are products of the eigenvectors of the
$h_j$s.  The eigenvectors can be labeled by a vector of single-body
eigenvalues, $\ket{\vec{\lambda}}\equiv
\ket{\lambda_{1},\ldots,\lambda_{n}}$.  The corresponding eigenvalues
of $h$ are given by the polynomial
\begin{equation}\label{eq:Pik}
  \Pi_k(\vec\lambda) =
  \bra{\vec\lambda} H \ket{\vec\lambda}
  =\bigg( \sum_{j=1}^n \lambda_j\bigg)^k
  = \sum^n_{a_1, \ldots , a_k}\lambda_{a_1}\cdots \lambda_{a_k}\;,
\end{equation}
which is symmetric under permutation of its arguments and is known
in the mathematical literature as the $k$th-degree elementary
symmetric polynomial (on $n$ variables).

To calculate the QCRB~\eqref{eq:semi-norm} for the $k$-body coupling
Hamiltonian~\eqref{eq:hs}, we have to calculate the maximum and
minimum eigenvalues of $H$ in terms of the eigenvalues of the
single-body operators $h_j$, the total number of constituents $n$,
and the degree of the coupling $k$.  Let $\lmax$ and $\lmin$ be the
largest and smallest eigenvalues of $h_j$.  We consider four cases.

\begin{enumerate}

\item $k$ odd.  The largest (smallest) eigenvalue of $H$ is
$\Lmax = (n\lmax)^k$ [$\Lmin=(n\lmin)^k$], corresponding to the
eigenvector $\ket{\Lmax}=\ket{\lmax,\ldots,\lmax}$ ($\ket{\Lmin} =
\ket{\lmin,\ldots, \lmin}$).

\item $k$ even, $\lmin\ge0$.  The same conclusions
apply as in Case~1.

\item $k$ even, $\lmax\le0$.  The same conclusions
apply as in Case~2, except that the roles of $\lmax$
and $\lmin$ are reversed: the largest (smallest) eigenvalue,
$\Lmax=(n\lmin)^k$ [$\Lmin=(n\lmax)^k$], corresponds to the
eigenvector that has every constituent in the state $\ket{\lmin}$
($\ket{\lmax}$).

\item $k$ even, $\lmin<0<\lmax$.  Since all the eigenvalues of
$H$ are nonnegative, the maximum eigenvalue is $\Lmax=(n\abslmax)^k$,
where $\abslmax\equiv\max\{|\lmax|,|\lmin|\}$, corresponding to all
the constituents being in either $\ket{\lmax}$ or $\ket{\lmin}$.  The
minimum eigenvalue comes from the string $\vec\lambda$, perhaps
containing all eigenvalues, that makes $\Pi_k(\vec\lambda)$ as close
to zero as possible.

\end{enumerate}

\noindent In Cases~1--3, the QCRB~\eqref{eq:semi-norm} takes the form
\begin{equation}
\label{eq:cr2}
\delta \gamma \geq \frac{1}{t n^k|\lmax^k - \lmin^k|}\;,
\end{equation}
displaying the $\cO(n^{-k})$ scaling found in~\cite{boixo07a}.

Case~4 requires further discussion regarding $\Lmin$.  We can bound
$\Lmin$ from above by considering strings $\vec\lambda$ that contain
only $\lmax$ and $\lmin$.  If $\lmax$ appears a fraction $p$ of the
time, the corresponding eigenvalue is
\begin{equation}
 \Pi_k(\vec\lambda) = [np\lmax + n(1-p)\lmin]^k\;.
\end{equation}
This eigenvalue can be minimized by making $p\lmax + (1-p)\lmin$ as
close to zero as possible, i.e., by choosing
\begin{equation}
\label{eq:p}
np = \bigg\lceil\frac{n|\lmin|}{\normh} \bigg\rfloor\;,
\end{equation}
where $\lceil x \rfloor$ denotes the nearest integer to $x$ and
$\normh=\lmax-\lmin$ is the semi-norm of the single-particle
operators.  The resulting eigenvalue can be written as
$\Pi_k(\vec\lambda)=(\delta\normh)^k$, where $\delta\le1/2$ is the
magnitude of the difference between $n|\lmin|/\normh$ and the integer
closest to it.  The minimum eigenvalue can be written as
\begin{equation}
\Lmin=[(\delta-\epsilon)\normh]^k\;,
\end{equation}
where $\epsilon$ ($0\le\epsilon\le1/2$) accounts for the fact that
strings containing other eigenvalues of the $h_j$s can generally make
$\Lmin$ smaller.  If the constituents are qubits, $\epsilon=0$.
Since $\normh/\abslmax\le2$, we have
\begin{equation}
\frac{\Lmin}{\Lmax}=
\(\frac{\delta-\epsilon}{n}\frac{\normh}{\abslmax}\)^k
\le n^{-k}\;.
\end{equation}

The upshot is that for even $k$ and $\lmin<0<\lmax$, the
QCRB~\eqref{eq:semi-norm} is given by
\begin{equation}
\delta\gamma\ge
\frac{1}{tn^k\abslmax^k}
\frac{1}{1-\Lmin/\Lmax}\;.
\label{eq:cr4}
\end{equation}
Thus the symmetric $k$-body coupling leads to a QCRB scaling as
$\cO(n^{-k})$ in all four cases.

A closely related $k$-body coupling Hamiltonian is the same as
Eq.~\eqref{eq:hs}, except that the self-interaction terms are
omitted, which might be more appropriate in some physical situations.
We analyze this alternative $k$-body coupling Hamiltonian in
App.~\ref{appA} and show that it also leads to a $\cO(n^{-k})$ QCRB
scaling when arbitrary input states are allowed.

Luis and collaborators~\cite{luis04a,beltran05a} showed that
$\cO(n^{-2})$ scalings can be achieved in principle using a Kerr-type
optical nonlinearity, and Luis generalized these results to optical
nonlinearities of arbitrary order in~\cite{luis07a}.
Reference~\cite{partner07a} proposes a method for synthesizing a
quadratic ($k=2$) Hamiltonian from a linear Hamiltonian by passing a
light beam twice through an atomic medium and finds a $\cO(n^{-2})$
scaling for this method.

\section{Attainable precision with pure product states} \label{sec3}

\subsection{General bound}
\label{S:exp}

The QCRB~\eqref{eq:semi-norm} gives the best possible measurement
precision, but can only be achieved for an optimal probe initial
state, i.e., one of the form
$(\ket{\Lmax}+e^{i\phi}\ket{\Lmin})/\sqrt2$, which the results of
Sec.~\ref{sec2} show is typically highly entangled.  In this section
we obtain lower bounds on $\delta\gamma$ in the situation where the
initial state is a pure product state,
\begin{equation}
\label{eq:prodstate}
\ket{\Psi_0} =
\ket{\psi_1}\otimes\cdots\otimes\ket{\psi_n}\;;
\end{equation}
for this purpose, we start from the state-dependent
QCRB~\eqref{eq:cramer-rao}.  Since all the one-body operators $h_j$
in the coupling Hamiltonian are assumed to be identical, it is
reasonable to expect that the optimal initial product state will have
all constituents in the same state, but we do not assume this at the
outset, instead allowing its moral equivalent to emerge from the
analysis.

The trick to evaluating $\Delta H$ is to partition the unrestricted
sum in Eq.~\eqref{eq:hs}, in which terms in the sum contain different
numbers of duplicate factors, into sums such that each term has the
same sort of duplicate factors.  Thus we write
\begin{align}
H=
&\sum_{(a_1,\ldots,a_k)}h_{a_1}\cdots h_{a_k}\nonumber\\
&+\binom{k}{2}\!\!\sum_{(a_1,\ldots,a_{k-1})}h_{a_1}\cdots h_{a_{k-2}}h_{a_{k-1}}^2
+\cdots\;,
\label{eq:hss}
\end{align}
where a summing range with parentheses, $(a_1,\ldots,a_l)$, denotes a
sum over all $l$-tuples that have no two elements equal.  The two
sums in Eq.~\eqref{eq:hss} are the leading- and subleading-order
terms in an expansion in which successive sums have fewer terms.  The
first sum in Eq.~\eqref{eq:hss}, in which the terms have no duplicate
factors, has $n!/(n-k)!=\cO(n^k)$ terms, and the second sum, in which
one factor is duplicated in each term, has $n!/(n-k-1)!=\cO(n^{k-1})$
terms.  The binomial coefficient multiplying the second sum accounts
for the number of ways of choosing the factor that is duplicated.
The next sums in the expansion, involving terms with factors $h_j^3$
and $h_j^2 h_l^2$, have $n!/(n-k-2)!=\cO(n^{k-2})$ terms.  These
expansions require that $n\ge k$, which we assume henceforth, and the
scalings we identify further require that $n\gg k$.

Given the expansion~\eqref{eq:hss}, the expectation value of $H$
has the form
\begin{align}
\avg{H}=
&\sum_{(a_1,\ldots,a_k)}\avg{h_{a_1}}\cdots\avg{h_{a_k}}\nonumber\\
&+\binom{k}{2}\!\!\sum_{(a_1,\ldots,a_{k-1})}
\avg{h_{a_1}}\cdots\avg{h_{a_{k-2}}}\avg{h_{a_{k-1}}^2}\nonumber\\
&+\cO(n^{k-2})\;.
\label{eq:expectHS}
\end{align}
The expression for $\avg{H^2}$ follows by replacing $k$ with
$2k$:
\begin{align}
\avg{H^2}=
&\sum_{(a_1,\ldots,a_{2k})}\avg{h_{a_1}}\cdots\avg{h_{a_{2k}}}\nonumber\\
&+\binom{2k}{2}\!\!\sum_{(a_1,\ldots,a_{2k-1})}
\avg{h_{a_1}}\cdots\avg{h_{a_{2k-2}}}\avg{h_{a_{2k-1}}^2}\nonumber\\
&+\cO(n^{2k-2})\;.
\label{eq:2k}
\end{align}
The rest of the analysis is based on an artful switching between
restricted and unrestricted sums.  By changing the initial sum in
Eq.~\eqref{eq:expectHS} to an unrestricted sum, we can rewrite
$\avg{H}$ to the required order as
\begin{align}
\avg{H}=
&\sum_{a_1,\ldots,a_k}\avg{h_{a_1}}\cdots\avg{h_{a_k}}\nonumber\\
&+\binom{k}{2}\!\!\sum_{(a_1,\ldots,a_{k-1})}
\avg{h_{a_1}}\cdots\avg{h_{a_{k-2}}}\Delta h_{a_{k-1}}^2\nonumber\\
&+\cO(n^{k-2})\;.
\label{eq:expectHS2}
\end{align}
Squaring this expression and changing the unrestricted sums back to
restricted ones, again keeping only the leading- and subleading-order
terms, gives
\begin{align}
\avg{H}^2=
&\sum_{(a_1,\ldots,a_{2k})}\avg{h_{a_1}}\cdots\avg{h_{a_{2k}}}\nonumber\\
&+\binom{2k}{2}\!\!\sum_{(a_1,\ldots,a_{2k-1})}
\avg{h_{a_1}}\cdots\avg{h_{a_{2k-2}}}\avg{h_{a_{2k-1}}}^2\nonumber\\
&+2\binom{k}{2}\!\!\sum_{(a_1,\ldots,a_{2k-1})}
\avg{h_{a_1}}\cdots\avg{h_{a_{2k-2}}}\Delta h_{a_{k-1}}^2\nonumber\\
&+\cO(n^{2k-2})\;.
\label{eq:expectHSsquare}
\end{align}

We can now find $(\Delta H)^2$ by subtracting Eq.~\eqref{eq:expectHSsquare}
from Eq.~\eqref{eq:2k}:
\begin{align}
\label{eq:varHS}
  (\Delta H)^2 &=
  k^2\!\!\!\sum_{(a_1,\ldots,a_{2k-1})}
  \avg{h_{a_1}}\cdots\avg{h_{a_{2k-2}}}\Delta h_{a_{k-1}}^2\nonumber\\
  &\phantom{=x}+\cO(n^{2k-2})\nonumber\\
  &=k^2\biggl(\sum_{j=1}^n\avg{h_j}\biggr)^{2(k-1)}
  \biggl(\sum_{j=1}^n\Delta h_j^2\biggr)+\cO(n^{2k-2})
  \;.
\end{align}
In the final form, we take advantage of the fact that in the now
leading-order sum, we can convert the restricted sum to an
unrestricted one.

To make the QCRB~\eqref{eq:cramer-rao} as small as possible, we need
to maximize the variance $(\Delta H)^2$ of Eq.~\eqref{eq:varHS}.  We
can immediately see that for fixed expectation values $\avg{h_j}$, we
should maximize the variances $\Delta h_j^2$, and this is done by
using for each constituent a state that lies in the subspace spanned
by $\ket{\lmax}$ and $\ket{\lmin}$.  Letting $p_j$ be the probability
associated with $\ket{\lmax}$ for the $j$th constituent, we have
\begin{align}
x_j\equiv\avg{h_j}&=p_j\lmax+(1-p_j)\lmin=\lmin+p_j\normh\;,\nonumber\\
\Delta h_j^2&=p_j\lmax^2+(1-p_j)\lmin^2 - x_j ^2\nonumber\\
&=\normh^2 p_j(1-p_j)\nonumber\\
&=(\lmax-x_j)(x_j-\lmin)\;.
\end{align}
Thus we should maximize
\begin{equation}
(\Delta H)^2=k^2
\biggl(\sum_{j=1}^n x_j\biggr)^{2(k-1)}\sum_{j=1}^n(\lmax-x_j)(x_j-\lmin)
\end{equation}
within the domain defined by $\lmin\le x_j\le\lmax$, $j=1,\ldots,n$.

Discarding potential extrema of $(\Delta H)^2$ given by $0=\sum_j
x_j$, since these either are minima or lie outside the relevant
domain, we find that the conditions for extrema of $(\Delta H)^2$
imply immediately that $x_j=x$ (and thus $p_j=p$) for $j=1,\ldots,n$.
Thus the optimal states in the initial product
state~\eqref{eq:prodstate} have the form
\begin{equation}
\ket{\psi_j}=\sqrt{p}\,\ket{\lmax}+e^{i\phi_j}\sqrt{1-p}\,\ket{\lmin}\;.
\label{eq:psij}
\end{equation}
The only possible difference between the states for different
constituents is in the relative phase between $\ket{\lmax}$ and
$\ket{\lmin}$.

Since the optimal constituent states live and evolve in a
two-dimensional subspace, we can regard the constituents effectively
as qubits, with standard basis states $\ket{0}=\ket{\lmax}$ and
$\ket{1}=\ket{\lmin}$, serving as the basis for constructing Pauli
operators $X$, $Y$, and $Z$.  Restricted to this subspace, the
operator $h$ takes the form
\begin{equation}
h=\lmax\ket{0}\!\bra{0}+\lmin\ket{1}\!\bra{1}=
\lambdabar\openone+\normh Z/2\;,
\end{equation}
where $\lambdabar\equiv(\lmax+\lmin)/2$ is the arithmetic mean of the
largest and smallest eigenvalues of $h$.

In the analyses in Secs.~\ref{sec4} and \ref{sec5}, we assume that
all the constituents have zero relative phase ($\phi_j=0$), giving an
initial state $\ket{\Psi_\beta}=\ket{\psi_\beta}^{\otimes n}$, where
\begin{align}
\ket{\psi_\beta}&=
e^{-i\beta Y/2}\ket{0}=\cos(\beta/2)\ket{0}+\sin(\beta/2)\ket{1}\;,\nonumber\\
p&=\cos^2(\beta/2)=(1+\cos\beta)/2\;.
\label{eq:psi}
\end{align}
Here we describe the one-body state $\ket{\psi_\beta}$ in terms of
the rotation angle $\beta$ about the $y$ axis that produces it from
$\ket{0}$.  The corresponding initial density operator is
\begin{equation}
\rho_\beta=\ket{\Psi_\beta}\!\bra{\Psi_\beta}=
\bigotimes_{j=1}^n
{1\over2}(\openone_j+X_j\sin\beta+Z_j\cos\beta)\;.
\label{eq:rhobeta}
\end{equation}

The variance of $H$ now takes the simple form
\begin{align}
(\Delta H)^2&=k^2 n^{2k-1}\avg{h}^{2(k-1)}(\Delta h)^2\nonumber\\
&=k^2 n^{2k-1}x^{2(k-1)}(\lmax-x)(x-\lmin)\;,
\label{eq:DeltaH}
\end{align}
which leads, in the QCRB~\eqref{eq:cramer-rao}, to a sensitivity that
scales as $1/n^{k-1/2}$ for input product states.  This should be
compared with the $\cO(n^{-k})$ scaling that can be obtained by using
initial entangled states~\cite{boixo07a}.  Notice that for $k\ge2$,
the $\cO(n^{-k+1/2})$ scaling is better than the $1/n$ scaling of the
Heisenberg limit, which is the best that can be achieved in the $k=1$
case even with entangled initial states.

The coupling Hamiltonian that has the self-interaction terms of
Eq.~(\ref{eq:hs}) removed is analyzed in App.~\ref{appA}.  We show
that for initial product states, this modified Hamiltonian has the
same leading-order behavior in the variance of $H$; it thus has
$\cO(n^{-k+1/2})$ scaling for initial product states and the same
optimal product states as we now find for the coupling
Hamiltonian~\eqref{eq:hs}.

\subsection{Optimal product states
\label{S:max}}

The problem of finding the optimal input product state is now reduced
to maximizing the $2k$-degree polynomial
\begin{align}
f(x)&\equiv x^{2(k-1)}(\lmax-x)(x-\lmin)\nonumber\\
&=x^{2(k-1)}\bigl(\normh^2/4-(x-\lambdabar)^2\bigr)
\end{align}
with respect to the single variable $x=\avg{h}$ on the domain
$\lmin\le x\le\lmax$.  The condition for an extremum is
\begin{align}
0=f'(x)
&=2x^{2k-3}\bigl[(k-1)\bigl(\normh^2/4-(x-\lambdabar)^2\bigr)\nonumber\\
&\phantom{=2x^{2k-3}\bigl[(k}-x(x-\lambdabar)\bigr]\;.
\label{eq:ext}
\end{align}
We assume $k\ge2$, because the $k=1$ case is already well understood.
For $k=1$, there is a single maximum at $x=\lambdabar$, corresponding
to equal probabilities for $\ket{\lmax}$ and $\ket{\lmin}$ and to
$(\Delta H)^2=n\normh^2/4$.

The polynomial $f$ vanishes at $x=\lmin$ and $x=\lmax$.  We can make
some general statements about the extrema of $f$ in three cases.

\begin{enumerate}
\item If $\lmin<0<\lmax$, $f$ has a minimum at $x=0$ and two
maxima within the allowed domain, one at a positive $x_+>\lambdabar$
and one at a negative $x_-<\lambdabar$.  The global maximum is at
$x_+$ ($x_-$) if $\abslmax=\lmax$ ($\abslmax=|\lmin|$).

\item If $\lmax>\lmin>0$, $f$ has a maximum at $x=0$, a minimum for
a positive $x_-<\lmin$, and a maximum within the allowed domain at
$x_+>\lambdabar$.  Only the last of these lies in the relevant domain.

\item If $\lmin<\lmax<0$, $f$ has a maximum at $x=0$, a
minimum for a negative $x_+>\lmax$, and a maximum within the allowed
domain at $x_-<\lambdabar$.  Only the last of these lies in the
relevant domain.
\end{enumerate}

These general observations are perhaps more enlightening than the
form of the (nonzero) solutions of Eq.~\eqref{eq:ext}:
\begin{equation}
x_\pm
=\biggl(1-\frac{1}{2k}\biggr)\lambdabar
\pm\frac{1}{2}
\sqrt{\frac{\lambdabar^2}{k^2}+
\biggl(1-\frac{1}{k}\biggr)\normh^2}\;.
\end{equation}
The $\pm$ here means the same thing as in the discussion of the three
cases above.

As $k$ increases, $x_+$ approaches $\lmax$, and $x_-$ approaches
$\lmin$.  Indeed, as $k\rightarrow\infty$, we have
$x_+=(1-1/2k)\lmax$, corresponding to $p_+=1-\lmax/2k\normh$ and
$(\Delta H)^2=(k/2e)(n\lmax)^{2k-1}\normh$, and $x_-=(1-1/2k)\lmin$,
corresponding to $p_-=-\lmin/2k\normh$ and $(\Delta H)^2 =
(k/2e)(-n\lmin)^{2k-1}\normh$.

An important limiting case, not covered in the discussion above,
occurs when $\lmin=-\lmax$.  Then the maxima occur symmetrically at
\begin{equation}
x_\pm=\pm\frac{1}{2}\normh\sqrt{1-1/k}\;,
\end{equation}
corresponding to probabilities
$p_\pm=\frac{1}{2}+x_\pm/\normh=\frac{1}{2}(1\pm\sqrt{1-1/k})=1-p_\mp$
and to $\sin\beta_\pm=\sqrt{1/k}\,$.  The two maxima lead to the same
variance,
\begin{equation}
(\Delta H)^2=k(1-1/k)^{k-1}n^{2k-1}(\normh/2)^{2k}\;,
\end{equation}
thus yielding a QCRB
\begin{equation}
\delta\gamma\ge\frac{2^{k-1}}{k^{1/2}(1-1/k)^{(k-1)/2}}
\frac{1}{t\,n^{k-1/2}\normh^k}\;.
\end{equation}
Of course, when $\lmin=-\lmax$, we can always choose units such that
$\lmax=1/2$ ($\normh=1$), which means that the single-body operators
are $h_j=Z_j/2$.  It is this situation that we analyze in the
remainder of this paper.

\section{Separable measurements \label{sec4}}

In the previous section we obtained the theoretical limits on the
measurement uncertainty with symmetric $k$-body couplings and initial
product states for the probe.  The theoretical bound is saturated by
a measurement of the so-called symmetric logarithmic
derivative~\cite{helstrom76a,holevo82a,braunstein94a,braunstein96a};
this measurement, in general, is entangled and depends on the value
of the parameter that we are attempting to estimate.  In this section
we show that for some Hamiltonians of interest, standard separable
measurements lead to uncertainties for small $\gamma$ that have the
same scaling as the theoretical bounds.  The restriction to small
values of $\gamma$ is not a strong limitation, because we can always
use feedback to operate in this regime, as we discuss in more detail
in Sec.~\ref{sec:mea}.

We consider the special case in which the single-body operators
are $h_j=Z_j/2$, leading to a coupling Hamiltonian
\begin{align}
  H = \bigg(\sum_j Z_j/2\bigg)^k = J_z^k\;.
\end{align}
Here we introduce $J_z$ as the $z$ component of a ``total angular
momentum'' corresponding to the effective qubits.  We assume an
initial state of the form~\eqref{eq:rhobeta}, and we let this state
evolve for a very short time, i.e., $\phi\equiv\gamma t\ll1$.  In the
remainder of the paper, we often work in terms of the parameter
$\phi$ instead of $\gamma$.  After the time evolution, we measure the
separable observable
\begin{equation}
J_y=\sum_j Y_j\;.
\end{equation}
Over $\nu$ trials, we estimate $\phi$ as a scaled arithmetic mean of
the results of the $J_y$ measurements.

The expectation value of any observable at time $t$ is given by
\begin{equation}
\label{mean}
\avg{M}_t = \Tr\(U^\dg MU\rho_\beta\)=\avg{U^\dag MU}\;,
\end{equation}
where $U = e^{-iH_\gamma t}=e^{-iH\phi}$, and where we introduce the
convention that an expectation value with no subscript is taken with
respect to the initial state.  For small $\phi$, we have
 \begin{equation}
 \label{comm}
U^\dg M U = M - i\phi[M,H] + \cO(\phi^2)\;.
 \end{equation}
Thus the expectation value and variance of $J_y$ at time $t$ take the
form
\begin{subequations}
\label{eq:Jystuff}
\begin{align}
\avg{J_y}_t&=\avg{J_y}-i\phi\avg{[J_y,H]}+\cO(\phi^2)\;,
\label{eq:avgJy}\\
(\Delta J_y)_t^2&=(\Delta J_y)_0^2\nonumber\\
&\phantom{=(}-i\phi
\bigl\langle(J_y-\avg{J_y})[J_y,H]+[J_y,H](J_y-\avg{J_y})\bigr\rangle\nonumber\\
&\phantom{=(}+\cO(\phi^2)\;.
\label{eq:varJy}
\end{align}
\end{subequations}

The initial expectation value and variance of $J_y$ are those of an
angular-momentum coherent state in the $x$-$z$ plane:
\begin{subequations}
\begin{align}
\avg{J_y}&=0\;,\\
(\Delta J_y)_0^2&=\avg{J_y^2}=
\frac{1}{4}\sum_{j,l}\avg{Y_jY_l}=
\frac{n}{4}\;.
\end{align}
\end{subequations}
In evaluating the other expectation values in Eqs.~\eqref{eq:Jystuff},
we can avail ourselves of the expansions used in Sec.~\ref{S:exp},
since we are only interested in the leading-order behavior in $n$.
To leading order, the coupling Hamiltonian has the form
\begin{align}
  H =
  \frac{1}{2^k}\sum_{(a_1,\ldots,a_k)} Z_{a_1}\cdots Z_{a_k}+\cO(n^{k-1})\;.
\end{align}
In this section we use $\approx\,$ to indicate equalities that are
good to leading order in $n$.  We can now write
\begin{align}
[J_y,H]&\approx\frac{1}{2^{k+1}}
\sum_{j=1}^n \sum_{(a_1,\ldots,a_k)}[Y_j,Z_{a_1}\cdots Z_{a_k}]\nonumber\\
&=\frac{i}{2^k}\sum_{l=1}^k\sum_{(a_1,\ldots,a_k)}
Z_{a_1}\cdots Z_{a_{l-1}}X_{a_l}Z_{a_{l+1}}\cdots Z_{a_k}\nonumber\\
&=\frac{ik}{2^k}\sum_{(a_1,\ldots,a_k)}X_{a_1}Z_{a_2}\cdots Z_{a_k}\;,
\end{align}
from which it follows that
\begin{align}
\avg{[J_y,H]}&\approx
\frac{ik}{2^k}\sum_{(a_1,\ldots,a_k)}\avg{X_{a_1}}\avg{Z_{a_2}}\cdots\avg{Z_{a_k}}\nonumber\\
&\approx ik\avg{J_x}\avg{J_z}^{k-1}\;.
\end{align}
Elaborating this procedure one step further, we can show that to
leading order in $n$, the expectation value in the second line of
Eq.~\eqref{eq:varJy} vanishes.  Our results to this point are
summarized by
\begin{subequations}
\begin{align}
\avg{J_y}_t&\approx\phi k\avg{J_x}\avg{J_z}^{k-1}+\cO(\phi^2)\nonumber\\
&=\phi k(n/2)^k\sin\beta\cos^{k-1}\!\beta+\cO(\phi^2)\;,\\
(\Delta J_y)_t&\approx\sqrt n/2+\cO(\phi^2)\;.
\end{align}
\end{subequations}

If we let our estimator $\phi_{\rm est}$ be the arithmetic mean of
the $\nu$ measurements of $J_y$, scaled by the factor
$(d\avg{J_y}_t/d\phi)^{-1}=1/k(n/2)^k\sin\beta\cos^{k-1}\!\beta$, we have
\begin{align}
\avg{\phi_{\rm est}}&=\frac{\avg{J_y}_t}{d\avg{J_y}_t/d\phi}
\approx\phi+\cO(\phi^2)\;,\\
\delta\phi
&\approx\frac{1}{\sqrt\nu}\frac{(\Delta J_y)_t}{|d\avg{J_y}_t/d\phi|}+\cO(\phi)\nonumber\\
&\approx\frac{1}{\sqrt\nu}\frac{2^{k-1}}{kn^{k-1/2}\sin\beta|\cos^{k-1}\!\beta|}+\cO(\phi)\;.
\end{align}
This scheme thus attains the $\cO(n^{-k+1/2})$ scaling that is the
best that can be achieved by initial product states.  In an analysis
of optical nonlinearities of arbitrary order, Luis~\cite{luis07a}
reported finding this $\cO(n^{-k+1/2})$ scaling.

The minimum of $\delta\phi$, occurring when $\sin\beta=\sqrt{1/k}$,
gives an optimal sensitivity
\begin{equation}
\delta\phi\approx
\frac{1}{\sqrt\nu}\frac{2^{k-1}}{k^{1/2}(1-1/k)^{(k-1)/2}}\frac{1}{n^{k-1/2}}+\cO(\phi)\;,
\end{equation}
which is identical to the optimal QCRB sensitivity for initial
product states.  For $k=2$, the case that is the subject of the next
section, the two optimal values of $\beta$ are $\beta=\pi/4$ and
$\beta=3\pi/4$, and the sensitivity becomes
\begin{equation}
\delta\phi\approx
\frac{1}{\sqrt\nu}\frac{2}{n^{3/2}}+\cO(\phi)\;.
\end{equation}

Aside from showing that the QCRB scaling for initial product states
can be achieved, the analysis in this section serves to illustrate
how the product-state scheme works in a regime that has a singularly
simple description. The $J_z^k$ coupling Hamiltonian induces a
nonlinear rotation about the $z$ axis, which rotates the probe
through an angle $\avg{J_y}_t/\avg{J_x}\approx \phi k\avg{J_z}^{k-1}$.
This rotation induces a signal in $J_y$ of size $\approx\phi
k\avg{J_x}\avg{J_z}^{k-1}$, which is $k\avg{J_z}^{k-1}$ times bigger
than for $k=1$, yet is to be detected against the same coherent-state
uncertainty $\sqrt n/2$ in $J_y$ as for $k=1$.  To take advantage of
the nonlinear rotation, we can't make the $J_x$ lever arm of the
rotation as large as possible, because the nonlinear rotation
vanishes when the initial coherent state lies in the equatorial
plane.  Nonetheless, we still win when we make the optimal compromise
between the nonlinear rotation and the lever arm. The optimal
compromise comes from maximizing $\avg{J_x}\avg{J_z}^{k-1}$, which
turns out to be exactly the same as finding the optimum in the QCRB
analysis of Sec.~\ref{S:max} because $\avg{X}=\sin\beta=\Delta Z$.

A more careful consideration of the terms neglected in this analysis
suggests that, as formulated in this section, the small-time
approximation requires that $\phi\ll 1/n^{k-1}$.  Nonetheless, the
analysis is consistent because $\phi$ can be resolved more finely
than this scale, i.e., $\delta\phi\,n^{k-1}=\cO(1/\sqrt n)$.  This
conclusion is confirmed by the more detailed analysis of the $k=2$
case in Sec.~\ref{sec5}.  On the other hand, the simple model of
coherent-state evolution, developed for $k=2$ in Sec.~\ref{sec5},
suggests the description of the preceding paragraph can be extended
to much larger times.  We return to this point in the Conclusion.

\section{Separable measurements for the interaction $H_\gamma=\gamma J_z^2$}
\label{sec5}

We focus now on the symmetric, $k=2$ coupling Hamiltonian
\begin{equation}
\label{eq:reyH}
H_\gamma=\gamma J_z^2=\gamma \biggl(\sum_j Z_j/2\biggr)^2\;.
\end{equation}
This is perhaps the most important example for practical applications
of nonlinear Hamiltonians to quantum metrology~\cite{boixo07a}, since
it occurs naturally whenever the strength of two-body interactions is
modulated by a parameter $\gamma$.  As suggested in~\cite{boixo07a},
one good place to look for this kind of coupling Hamiltonian is in
Bose-Einstein condensates.  Indeed, in~\cite{choi07a,rey07a}, it is
shown how this Hamiltonian can be implemented using the internal
atomic states of BECs.  In analyzing the BEC scenario,
Ref.~\cite{rey07a} finds a sensitivity that scales as $\cO(1/n)$ for
separable measurements made on a probe that evolves from an an
initial product state chosen to be an angular-momentum coherent state
in the equatorial plane.  The results in the previous sections show
that we should be able to improve this scaling to $\cO(n^{-3/2})$
through a wiser choice of the initial coherent state.  In this
section we analyze this situation in some detail.

We take the initial state of the $n$-qubit probe to be an
angular-momentum coherent state that is at an angle $\beta$ from the
$z$ axis in the $x$-$z$ plane.  This state is obtained from the
coherent state along the $z$ axis, $|J,J\rangle=\ket{0}^{\otimes n}$,
by a rotation through $\beta$ about the $y$ axis:
\begin{equation}
 \label{eq:is}
    \ket{\Psi_\beta} =
    e^{-i \beta J_y} \ket{J,J} =
    (e^{-i\beta Y/2}\ket{0})^{\otimes n}\;.
\end{equation}
The rotation about the $y$ axis and the nonlinear rotation under the
interaction Hamiltonian~\eqref{eq:reyH} both leave the state in the
$(2J+1)$-dimensional subspace with angular momentum $J=n/2$, so we
can use the basis $\ket{J,m}$ of $J_z$ eigenstates for this subspace,
with $m=-J,\ldots,J$.  The initial probe state used in~\cite{rey07a}
is a special case, $\beta=-\pi/2$.

The state $\ket{\Psi_\beta}=\sum_{m=-J}^J d_m\ket{J,m}$, can be
expanded in the basis $\ket{J,m}$ using a reduced Wigner rotation
matrix~\cite{sakurai94a}
\begin{align}
\label{eq:djm}
    d_m&\equiv d^J_{mJ}(\beta)=\bra{J,m}e^{-i\beta J_y}\ket{J,J}\nonumber\\
    &=\sqrt{\frac{(2J)!}{(J+m)!(J-m)!}}
    [\cos(\beta/2)]^{J+m}[\sin(\beta/2)]^{J-m}\;.
\end{align}
At time $t=\phi/\gamma$, the state of the probe becomes
\begin{equation}\label{eq:ts}
\ket{\Psi_\beta(t)} = e^{-i\phi J_z^2}\ket{\Psi_\beta}=
\sum_{m=-J}^J d_m e^{-i\phi m^2}\ket{J,m}\;.
\end{equation}

\subsection{Measurements}\label{sec:mea}

We now look at the attainable measurement uncertainties using both
$J_x = \sum X_j/2$ and $J_y =\sum_j Y_j/2$ measurements on the final
state of the probe.  It turns out that $J_x$ and $J_y$ measurements
are on nearly the same footing, with $J_y$ measurements being
marginally better, for all $\beta$ except $\beta=\pi/2$.  For very
short times, the superiority of $J_y$ measurements for
$\beta\ne\pi/2$ is clear from the analysis in Sec.~\ref{sec4}, since
the change in $\langle J_y \rangle$ is linear in $\phi$, whereas the
change in $\langle J_x \rangle$ is quadratic.  What happens for
longer times and for $\beta=\pi/2$ cannot be addressed by the
short-time analysis in Sec.~\ref{sec3}.  What we find in this section
is that both $J_x$ and $J_y$ measurements can achieve the optimal
scaling obtained in Sec.~\ref{sec3} .  For $\beta=\pi/2$, $J_y$
measurements provide no information about $\phi$, but $J_x$
measurements achieve the $\cO(n^{-1})$ scaling found
in~\cite{rey07a}.  These conclusions assume no decoherence, and in
Sec.~\ref{sec:dec} we explore the impact of decoherence on the
ability to achieve super-Heisenberg scalings with the symmetric,
$k=2$ coupling Hamiltonian.

For measurements of $J_x$ or $J_y$, the sensitivity is given by
\begin{equation}
\label{eq:exaccuracy}
\delta\phi_{x,y} = t\,\delta\gamma_{x,y} =
\frac{(\Delta J_{x,y})_\phi}{|d\avg{J_{x,y}}_\phi/d\phi|}
\end{equation}
(for this subsection, we revert to our practice of omitting the
$1/\sqrt\nu$ statistical factor from our sensitivity formulas).

The expressions needed to calculate $\delta\phi$ for $J_x$ and $J_y$
measurements are derived in Appendix~\ref{appB}.  These results are
conveniently expressed in terms of the raising and lowering operators
\begin{equation}
J_\pm=J_x\pm iJ_y\;,
\end{equation}
since we can write
\begin{align}
\label{eq:JxandJy}
&\avg{J_x}_\phi=\mathrm{Re}\big(\avg{J_+}_\phi\big)\;,
\qquad
\avg{J_y}_\phi=\mathrm{Im}\big(\avg{J_+}_\phi\big)\;,\\
&\avg{J_{x,y}^2}_\phi=
\frac{1}{4}\avg{J_+ J_- + J_- J_+}_\phi
\pm\frac{1}{2}\mathrm{Re}\big(\avg{J_+^2}_\phi\big)\;,
\label{eq:Jxysq}
\end{align}
where the upper sign in Eq.~\eqref{eq:Jxysq} applies to
$J_x$ and the lower sign to $J_y$.  In Appendix~\ref{appB},
we show that
\begin{align}
\label{eq:Jplus}
&\avg{J_+}_\phi=
J\sin\beta(\cos\phi+i\sin\phi\cos\beta)^{2J-1}\nonumber\\
&\phantom{\avg{J_+}_\phi\,}=J\sin\beta\,r^{2J-1}e^{i(2J-1)\theta}\;,\\
\label{eq:JplusJminus}
&\frac{1}{2}\avg{J_+ J_- + J_- J_+}_\phi
=J+\frac{J(2J-1)}{2}\sin^2\!\beta\;,\\
\label{eq:Jplussq}
&\avg{J_+^2}_\phi
=\frac{J(2J-1)}{2}\sin^2\!\beta(\cos2\phi+i\sin2\phi\cos\beta)^{2(J-1)}\nonumber\\
&\phantom{\avg{J_+^2}_\phi\,}=\frac{J(2J-1)}{2}\sin^2\!\beta\,R^{2(J-1)}e^{2i(J-1)\Theta}\;,
\end{align}
where
\begin{subequations}
\label{eq:rt}
\begin{align}
\label{eq:r}
r&=(1-\sin^2\!\phi\sin^2\!\beta)^{1/2}\;,\\
\theta&=\tan^{-1}(\tan\phi\cos\beta)\;,
\label{eq:t}
\end{align}
\end{subequations}
and
\begin{subequations}
\label{eq:RT}
\begin{align}\label{eq:R}
R&=(1-\sin^2\!2\phi\sin^2\!\beta)^{1/2}\;,\\
\Theta&=\tan^{-1}(\tan2\phi\cos\beta)\;.
\label{eq:T}
\end{align}
\end{subequations}
Plugging these results into Eqs.~\eqref{eq:JxandJy}
and~\eqref{eq:Jxysq}, we arrive at
\begin{align}
 \label{eq:Jx}
 \avg{J_x}_\phi &= J \sin\beta\,r^{2J-1}\cos[(2J-1)\theta]\;,\\
 \avg{J_y}_\phi &= J \sin\beta\,r^{2J-1}\sin[(2J-1)\theta]\;,
 \label{eq:Jy}
\end{align}
and
\begin{align}
\avg{J_{x,y}^2}_\phi&=
\frac{J}{2}
+\frac{J(2J-1)}{4}\sin^2\!\beta\nonumber\\
&\phantom{=\frac{J}{2}+x}\times\big(1\pm R^{2(J-1)}\cos[2(J-1)\Theta]\big)\;.
\label{eq:Jxysq2}
\end{align}
In using these results in what follows, it is easier to deal directly
with the first forms in Eqs.~\eqref{eq:Jplus} and~\eqref{eq:Jplussq}
rather than working with the functions $r$, $\theta$, $R$, and
$\Theta$.

The expectation values $\avg{J_{x,y}}_\phi$ change sign when $\phi$
advances by $\pi$.  This means that their squares and absolute
values, which are all that appear in the
sensitivity~\eqref{eq:exaccuracy}, are periodic with period $\pi$.
The second moments $\avg{J_{x,y}^2}_\phi$ are periodic with period
$\pi/2$.  The upshot is that the uncertainties $\Delta J_{x,y}$ and
the precision $\delta\phi_{x,y}$ are periodic with period $\pi$. This
$\pi$-periodicity is a consequence of periodic revivals in the
evolved state $\ket{\Psi_\beta(t)}$.

\begin{figure*}[!ht]
\begin{center}
\resizebox{8 cm}{4.8 cm}{\includegraphics{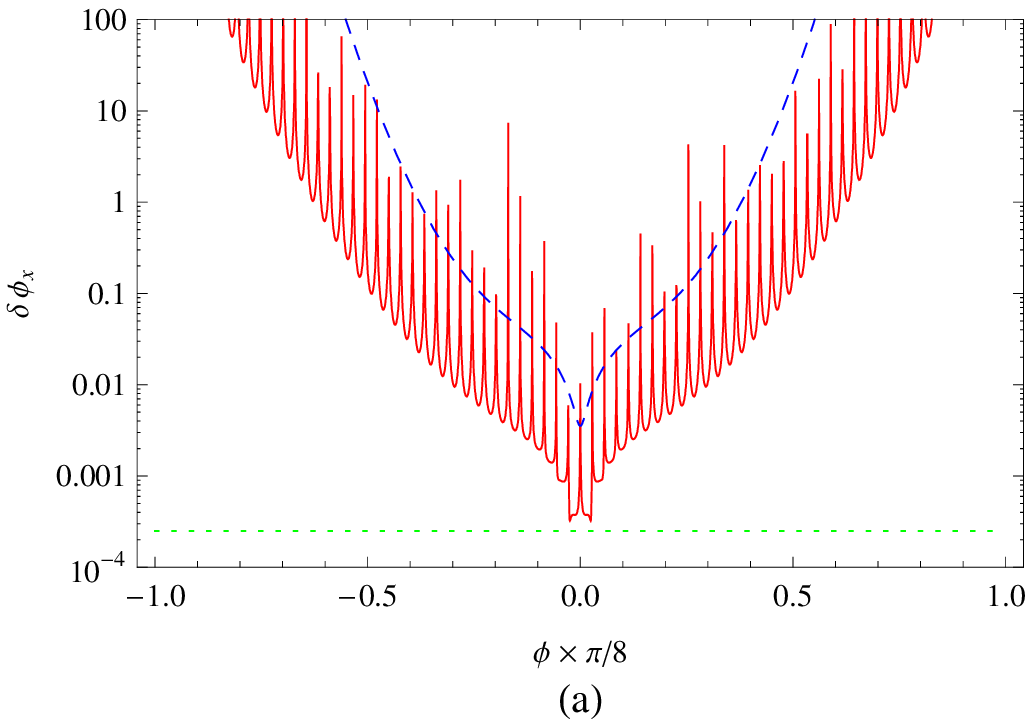}}\hspace{3em}
\resizebox{8 cm}{4.8 cm}{\includegraphics{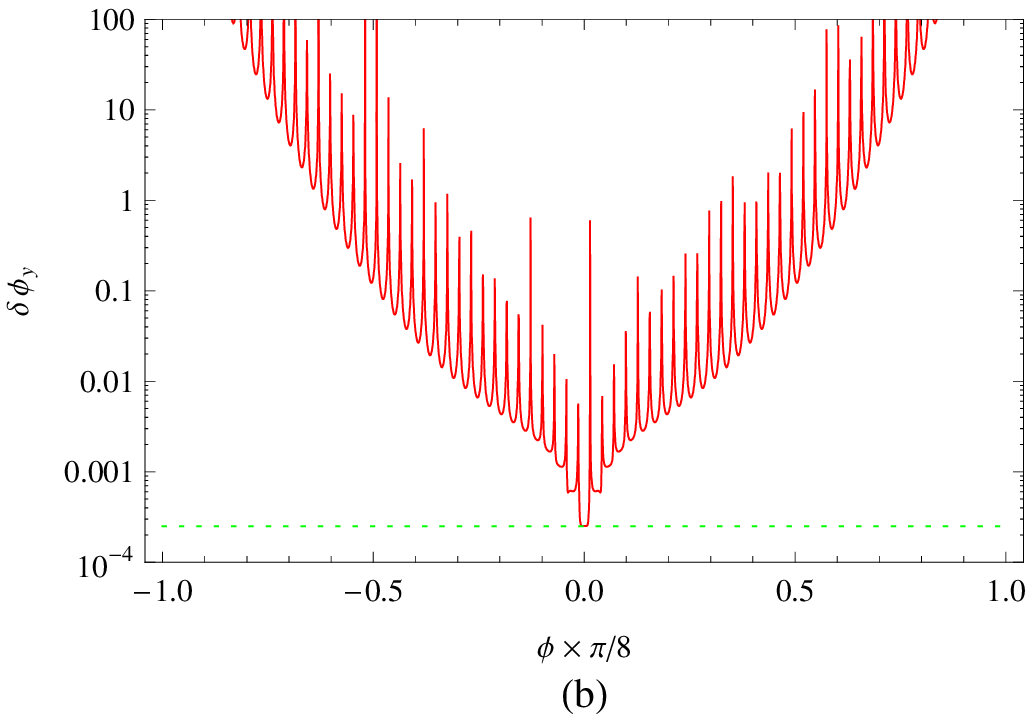}}
\caption{(Color online) Sensitivity (solid red lines) vs.~$\phi$
($-\pi/8\le\phi\le\pi/8$) using an optimal initial state at angle
$\beta=\pi/4$: (a)~$J_x$ measurements; (b)~$J_y$ measurements.  The
total angular momentum $J$ of the probe is 200, corresponding to
$n=400$.  The lower bound on the sensitivity, $1/\sqrt2 J^{3/2}$, is
plotted as the dotted (green) line.  The sensitivity is characterized
by rapidly oscillating fringes and a decay of sensitivity away from
the best sensitivities near $\phi=0$.  The sensitivity patterns
repeat with periodicity $\pi$; only a quarter of a period is plotted
because the sensitivity worsens even more outside the plotted region.
Part~(a) also shows the sensitivity for $J_x$ measurements when
$\beta=\pi/2$ (dashed blue line); notice the absence of fringes in
this case and the substantially degraded sensitivity.}
\label{fig:sensitivities}
\end{center}
\end{figure*}

The main features of the sensitivity $\delta\phi$ for measurements of
$J_x$ and $J_y$ can be gleaned from Fig.~\ref{fig:sensitivities}.  It is clear
from these plots that the best sensitivity is achieved when $\phi$ is
near zero and also, because of the periodicity of $\delta\phi$, when
$\phi$ is near $q\pi$, for $q$ any integer.

When $J$ is large, we can develop a good approximation for the entire
region of high sensitivity, where $\phi$ is small, by writing
\begin{subequations}
\label{eq:mainapprox}
\begin{align}
&(\cos\phi+i\sin\phi\cos\beta)^{2J-1}\simeq e^{2iJ\phi\cos\beta}e^{-J\phi^2\sin^2\!\beta}\;,\\
&(\cos2\phi+i\sin2\phi\cos\beta)^{2(J-1)}\simeq e^{4iJ\phi\cos\beta}e^{-4J\phi^2\sin^2\!\beta}\;.
\end{align}
\end{subequations}
These approximations are good to second order in $\phi$ in the
exponent.  When $\phi$ is near $q\pi$, the same approximations can be
had by replacing $\phi$ with $\phi-q\pi$.  The complex exponentials
give rise to rapidly oscillating fringes in $\avg{J_{x,y}}_\phi$ and
$\avg{J_{x,y}^2}_\phi$, with periods $\sim 1/J\cos\beta$; the slower
Gaussian envelopes take these expressions to zero when $\phi$ is a few
times $|\sin\beta|/\sqrt J$.

It is not hard to work out the sensitivity in this approximation, but
the formulas are sufficiently messy that they are little more
illuminating than the exact expressions.  We can, however, develop a
very simple, yet instructive picture of the fringes by keeping them,
but assuming that $\phi$ is small enough that the Gaussian envelopes
have yet to become effective, i.e., $\sqrt J\phi\sin\beta$ is
somewhat smaller than 1.  In this approximation, the fringes are
uniform in $\phi$, and we obtain
\begin{align}
\avg{J_x}_\phi&\simeq J\sin\beta\cos(2J\phi\cos\beta)\;,\\
\avg{J_y}_\phi&\simeq J\sin\beta\sin(2J\phi\cos\beta)\;,\\
(\Delta J_x)_\phi^2&\simeq\frac{J}{2}[1-\sin^2\!\beta\cos^2(2J\phi\cos\beta)]\;,\\
(\Delta J_y)_\phi^2&\simeq\frac{J}{2}[1-\sin^2\!\beta\sin^2(2J\phi\cos\beta)]\;.
\label{eq:DeltaJy}
\end{align}
These lead to sensitivities
\begin{align}
\label{eq:xapp}
\delta\phi_x^2&\simeq
\frac{1}{2J^3}
\frac{1-\sin^2\!\beta\cos^2(2J\phi\cos\beta)}{\sin^2\!2\beta\sin^2(2J\phi\cos\beta)}\;,\\
\delta\phi_y^2&\simeq
\frac{1}{2J^3}
\frac{1-\sin^2\!\beta\sin^2(2J\phi\cos\beta)}{\sin^2\!2\beta\cos^2(2J\phi\cos\beta)}\;.
\label{eq:yapp}
\end{align}

\begin{figure*}[!ht]
\begin{center}
\resizebox{8 cm}{4.8 cm}{\includegraphics{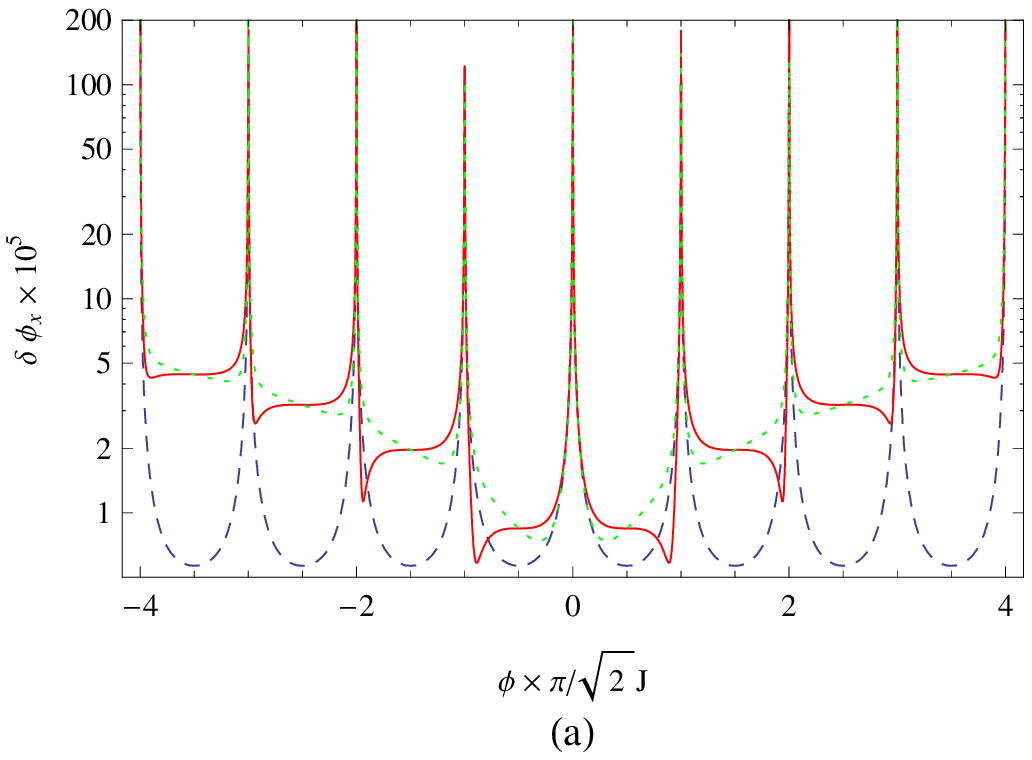}}\hspace{3em}
\resizebox{8 cm}{4.8 cm}{\includegraphics{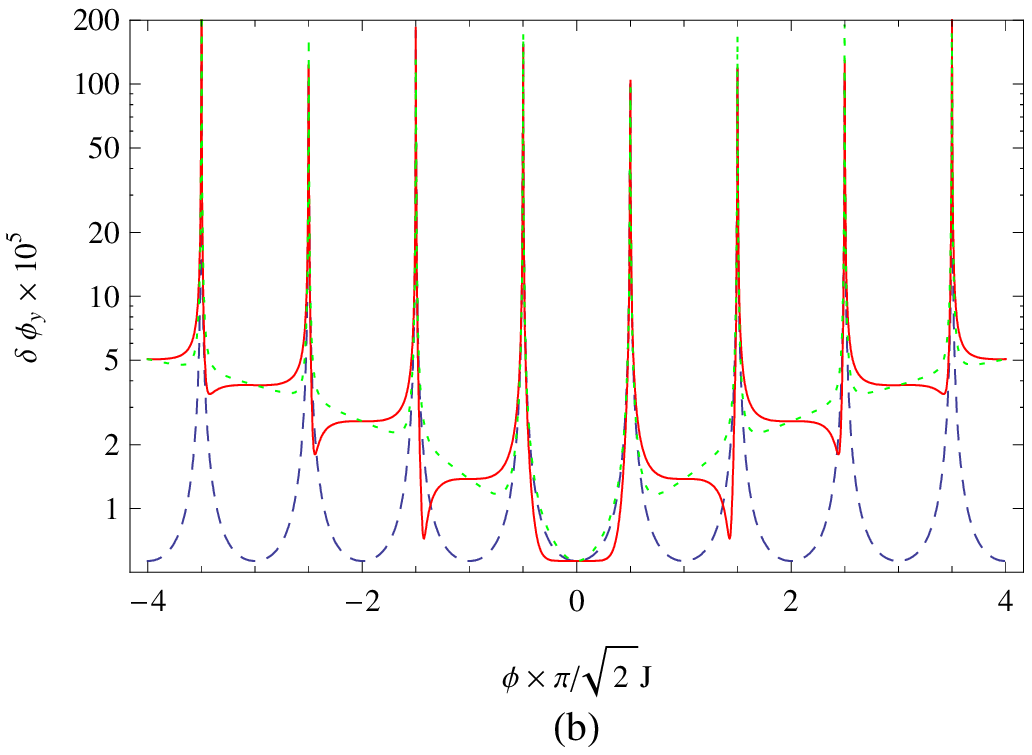}} \caption{(Color
online) Central few fringes of the measurement precision for
$\beta=\pi/4$ and $J=2\,500$ ($n=5\,000$): (a)~$J_x$ measurements;
(b)~$J_y$ measurements.  The solid (red) lines are the exact
sensitivities, the dashed (blue) lines are the sensitivities given by
the uniform-fringe approximation of Eqs.~\protect{\eqref{eq:xapp}}
and \protect{\eqref{eq:yapp}},\hspace{30em}
\protect{\vspace{.2em}}
\centerline{ $\displaystyle{
\delta\phi_x\simeq\frac{1}{2J^{3/2}}\sqrt{1+\frac{1}{\sin^2(2J\phi\cos\beta)}}\;,
\qquad
\delta\phi_y\simeq\frac{1}{2J^{3/2}}\sqrt{1+\frac{1}{\cos^2(2J\phi\cos\beta)}}}\;,$}
\noindent and the dotted (green) lines are the Gaussian-envelope
approximation of Eqs.~\protect{\eqref{eq:mainapprox}}.  The
uniform-fringe approximation locates the fringes precisely, but
misses entirely the degradation in sensitivity as one moves away from
the central fringes and also fails to characterize accurately the
shape of the fringes. The Gaussian-envelope approximation improves on
this performance by capturing the degradation of sensitivity quite
well, but still fails on the fringe shapes.  Even the central fringe
for $J_y$ measurements is noticeably flatter than in the two
approximations.  To get the best sensitivity, one should operate
right on the central fringe, at $\phi=q\pi$, for $J_y$ measurements
and on one of the two central fringes, centered at
$\phi=q\pi\pm\pi/4J\cos\beta$ for $J_x$ measurements.  Notice that
$J_x$ measurements achieve nearly optimal sensitivity at points near
the outside of these two central fringes.} \label{fig:fringes}
\end{center}
\end{figure*}

Within this uniform-fringe approximation, the best sensitivities are
achieved at the troughs of the fringes: the best operating points
are, for $J_x$ measurements,
\begin{equation}
\phi=q\pi+\frac{(s+1/2)\pi}{2J\cos\beta}\;,
\end{equation}
and for $J_y$ measurements,
\begin{equation}
\phi=q\pi+\frac{s\pi}{2J\cos\beta}\;,
\end{equation}
where $q$ and $s$ are integers.  At these operating points, the
sensitivity for both measurements becomes
\begin{equation}
\delta\phi_{x,y}\simeq\frac{1}{\sqrt 2 J^{3/2}|\sin2\beta|}\;,
\end{equation}
which takes on its optimal value, $1/\sqrt2J^{3/2}=2/n^{3/2}$, when
$\beta=\pi/4$ or $\beta=3\pi/4$, in agreement with the analyses in
Secs.~\ref{sec3} and \ref{sec4}.  This $\cO(J^{-3/2})$ scaling has
been found independently by Choi and Sundaram~\cite{choi07a} and for
nonlinear optical systems by Luis and
collaborators~\cite{luis04a,beltran05a}. For $\beta=\pi/4$ and
$J=2\,500$, Fig.~\ref{fig:fringes} plots the central fringes of the
approximate sensitivities~\eqref{eq:xapp} and \eqref{eq:yapp} and
compares them with the exact sensitivities and the Gaussian
approximation for measurements of $J_x$ and $J_y$.

As we show in App.~\ref{appB}, within the uniform-fringe
approximation, the evolved state~\eqref{eq:ts} is an angular-momentum
coherent state that makes an angle $\beta$ with the $z$ axis and that
rotates around the $z$ axis with angular velocity $2\gamma
J\cos\beta$.  The enhanced sensitivity available from a quadratic
Hamiltonian is a consequence of this increased rotation rate, which
is greater by a factor of $2J\cos\beta=2\avg{J_z}$ than that
available from a linear Hamiltonian.  This same conclusion came out
of the short-time analysis of Sec.~\ref{sec4}, but it is stronger now
because the uniform-fringe approximation is much better than the
short-time approximation.  The short-time approximation requires that
$J\phi\ll1$ and thus describes correctly only the center of the
central fringe for $J_y$ measurements.  In contrast, the
uniform-fringe approximation only requires that $\phi\sqrt
J|\sin\beta|\ll1$; within this requirement, there can be several
fringes, i.e., $2J\phi\cos\beta$ can be somewhat larger than $\pi$,
provided that $J\gg\tan^2\beta$.  The more accurate uniform-fringe
approximation allows us to see the other near-optimal operating
points for $J_y$ measurements and to see the optimal operating points
for $J_x$ measurements, which lie not at $\phi=0$, but at
$\phi=\pm\pi/4J\cos\beta$.  As $\beta$ approaches $\pi/2$, the
fringes become wider and wider, making the uniform-fringe
approximation reliable only for larger and larger values of $J$.  For
$\beta=\pi/2$, the fringes disappear entirely, and a separate
analysis is required to find the scaling for $J_x$ measurements
(since $\avg{J_y}_\phi=0$ for $\beta=\pi/2$, measurements of $J_y$
provide no information about~$\phi$).

That the final state~\eqref{eq:ts}, within the region of high
sensitivity, is approximately an angular-momentum coherent state
tells us two important things.  First, even though the quadratic
Hamiltonian will generate entanglement from a product state, this
entanglement plays no role in the enhanced sensitivity.  The improved
sensitivity comes from the increased rotation rate of the coherent
state, which is a product state, having no entanglement among the
probe constituents.  Indeed, for the measurements we consider here,
the deviation from being a coherent state makes the sensitivity
worse.  Second, that the probe state is approximately a product state
within the region of high sensitivity hints that this scheme should
not be as fragile in the presence of decoherence as schemes that rely
on initial entanglement. We investigate the impact of decoherence in
Sec.~\ref{sec:dec} and show that the $\cO(n^{-3/2})$ scaling is
unaffected by phase decoherence.

To achieve the optimal sensitivity for $J_x$ or $J_y$ measurements,
we need to operate within the appropriate central fringe, of width
$\pi/\sqrt 2J$ for $\beta=\pi/4$.  This can be done by using an
adaptive feedback procedure, which we discuss in the context of $J_y$
measurements.  The feedback procedure is carried out in several
steps, in each of which the quantity that is estimated is $\phi -
\phi_{\rm est}$, where $\phi_{\rm est}$ is the estimate of $\phi$
from the previous step.  At each step, we choose $J=n/2$ so that
$\phi-\phi_{\rm est}$ is with very high probability close to the
center of the central fringe, and we use $\nu$ probes to determine
$\phi$ with greater precision for the next step.  As we obtain
progressively refined estimates of $\phi$, the quantity being
estimated becomes smaller and smaller, always lying well within a
sequence of progressively finer central fringes.

To check that this procedure works and to determine its scaling
properties, imagine that we determine $\phi/2\pi$ bit by bit.
At step~$l$, we determine the $l$th bit of $\phi/2\pi$ by choosing
$J=J_l$ so that the precision is given by
\begin{equation}
\frac{\delta\phi_l}{2\pi}=\frac{1}{\sqrt\nu}\frac{1}{2\pi\sqrt 2 J_l^{3/2}}
=\frac{1}{f2^l}\;,
\end{equation}
where the factor $f\sim\mbox{3--10}$ is chosen to ensure that we get
the right $l$th bit with very high probability.  This gives
\begin{equation}
J_l=\frac{1}{2\nu^{1/3}}\left(\frac{f2^l}{\pi}\right)^{2/3}\;.
\end{equation}
We must, of course, choose $J_l$ to be an integer or half-integer, so
we choose the nearest one, but this detail does not change the
resource calculation significantly, so we ignore it. At step~$l+1$,
$\phi-\phi_{\rm est}$ lies well within the central fringe, as we see
from
\begin{equation}
\frac{\delta\phi_l}{\pi/\sqrt 2 J_{l+1}}=
\frac{2^{7/6}}{\pi\nu^{1/3}}\left(\frac{\pi}{f2^l}\right)^{1/3}\;.
\end{equation}
Indeed, because of the $\cO(J^{-3/2})$ scaling, the quantity being
estimated is buried progressively deeper fractionally in the central
fringe as we step through the procedure, despite the fact that the
central fringe is itself narrowing exponentially.

Suppose now that we use this procedure to estimate $L$ bits of $\phi$.
The total number of constituents used,
\begin{equation}
N=\nu\sum_{l=1}^L 2J_l=\left(\frac{2\nu f}{\pi}\right)^{2/3}
\frac{2^{2L/3}-1}{2^{2/3}-1}\;,
\end{equation}
is dominated by the last step, as is typical in these feedback
procedures.  The ultimate precision displays the $\cO(N^{-3/2})$
scaling,
\begin{align}
2\pi 2^{-L}&=\frac{4f}{(2^{2/3}-1)^{3/2}}\frac{\nu}{N^{3/2}}\nonumber\\
&=\frac{2f}{(2^{2/3}-1)^{3/2}}\frac{1}{\sqrt\nu}\frac{2}{(N/\nu)^{3/2}}\;,
\end{align}
with a small additional overhead given by the factor
$2f/(2^{2/3}-1)^{3/2}$.

As $\beta$ becomes smaller, the uniform-fringe approximation becomes
progressively better, since the fringes oscillate rapidly and the
Gaussian envelopes become very broad. On the other hand, the signal
in $J_x$ and $J_y$ disappears, making the sensitivity worsen as
$1/\sin^2\!2\beta$.

At the other extreme, as $\beta$ approaches $\pi/2$, the
uniform-fringe approximation becomes poorer as the fringes become as
wide as the Gaussian envelopes and loses validity entirely when
$J|\cot\beta|\sim1$.  When $\beta = \pi/2$, which is the initial
probe state analyzed by Rey~{\em et~al.}~\cite{rey07a}
(Ref.~\cite{rey07a} actually uses $\beta=-\pi/2$, but this state is
equivalent to $\beta=\pi/2$ for purposes of these measurements),
$\ket{J_y}_\phi=0$, making $J_y$ measurements useless for extracting
information about $\phi$.  Thus we have to choose the $J_x$
measurement.  The dashed line in Fig.~\ref{fig:sensitivities} shows
that the optimal operating point is $\phi=0$ (or, more generally,
$\phi=q\pi$).  Near $\phi=0$, the expectation value and variance of
$J_x$ are given by
\begin{align}
\avg{J_x}_\phi\simeq J-\frac{J(2J-1)}{2}\phi^2\;,\\
(\Delta J_x)^2_\phi\simeq J(2J-1)\phi^2\;,
\end{align}
where the approximations hold for $\phi\ll1/\sqrt J$.  The resulting
optimal sensitivity,
\begin{equation}
\delta\phi=\frac{(\Delta J_x)_\phi}{|d\avg{J_x}_\phi/d\phi|}\simeq
\frac{1}{\sqrt{J(2J-1)}}\;,
\end{equation}
has the $\cO(J^{-1})$ sensitivity scaling found in~\cite{rey07a}.

We can relate these results to the general lower-bound analysis in
Sec.~\ref{sec3} by noting that $\beta = \pi/2$ means that $\avg{h}
=\avg{Z}/2=0$.  This means that the dominant sums in the expansions
of Eqs.~\eqref{eq:2k} and \eqref{eq:expectHSsquare} are those that
contain only squares of $h_j$s.  The number of terms in these sums
scales as $\cO(n^k)$, which yields a sensitivity that scales as
$\cO(n^{-k/2})$.

\begin{figure}[!ht]
\begin{center}
\resizebox{7.5 cm}{4.8 cm}{\includegraphics{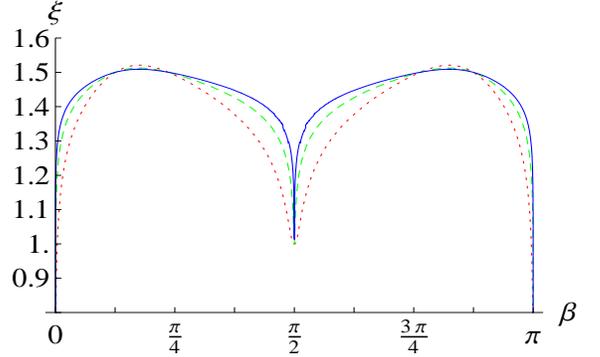}}
\caption{(Color online) Scaling exponent $\xi$ for $J_x$
measurements. The dotted (red) line is for $J=10^3$, the dashed
(green) line for $J=10^5$, and the solid (blue) line for $J = 10^7$.}
\label{fig:jxexp}
\end{center}
\end{figure}

\begin{figure}[!ht]
\begin{center}
\resizebox{7.5 cm}{4.8 cm}{\includegraphics{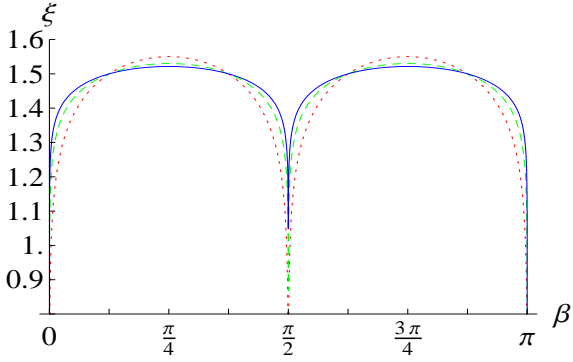}}
 \caption{(Color online) Scaling exponent $\xi$ for $J_y$
measurements. The dotted (red) line is for $J=10^3$, the dashed
(green) line for $J=10^5$, and the solid (blue) line for $J = 10^7$.}
\label{fig:jyexp}
\end{center}
\end{figure}

To gain further insight into the scaling behavior, we plot the scaling
exponent $\xi$ in $\delta \phi=\cO( n^{-\xi})$ as a function of
$\beta$ for $J_x$ measurements (Fig.~\ref{fig:jxexp}) and $J_y$
measurements (Fig.~\ref{fig:jyexp}), using three very large values of
$J$.  For $J_y$ measurements we calculate $\xi$ at the optimal
operating point, $\phi=0$.  For $J_x$ measurements, the optimal
operating point is a function of $\beta$, but a good compromise
point, which works well over the entire range of $\beta$, is
$1/\sqrt2J$, so we calculate the scaling exponent at this point for
all values of $\beta$.  An investigation of nearby operating points
scaling as $1/J$ gives plots with no discernible differences for the
large values of $J$ under consideration.  The main differences
between $J_x$ and $J_y$ measurements are the following: (i)~right at
$\beta=\pi/2$, $J_x$ measurements have a scaling exponent of 1,
whereas $J_y$ measurements provide no information about $\gamma$;
(ii)~for $J_y$ measurements, the plot of scaling exponent has two
humps, nearly symmetric about $\beta=\pi/4$ and $\beta=3\pi/4$,
whereas for $J_x$ measurements, the scaling exponent is better on the
outside of the humps.  The overall trend is for both measurements to
have a scaling exponent of $\xi=3/2$ in the limit of large $J$,
except at $\beta=0$, $\pi/2$, and $\pi$.

\subsection{Decoherence}\label{sec:dec}

The coherent-state model suggests that our generalized quantum
metrology scheme with initial product states should not display the
fragility of entangled protocols in the presence of decoherence.  We
can investigate this possibility by considering independent dephasing
of the effective qubits, described by the Lindblad equation
\begin{equation}
  \dot \rho = -\frac{\Gamma}{2}(Z\rho Z -\rho)\;,
\end{equation}
where $\tau_2=\Gamma^{-1}$ is the dephasing time.  Since dephasing
commutes with the quadratic Hamiltonian, we can shunt its effects to
the final time $t$, whence it maps the Pauli operators of each
effective qubit in the following way:
\begin{subequations}
\label{eq:decmap}
\begin{align}
  X & \rightarrow e^{-\Gamma t} X \;,\\
  Y & \rightarrow e^{-\Gamma t} Y \;,\\
  Z & \rightarrow Z\;.
\end{align}
\end{subequations}
To obtain the effect of the decoherence on the expectation values and
variances of the measured operators at the time of measurement, it is
easiest to use the adjoint map~\cite{huelga97a}, which for this
simple case is identical to the map~\eqref{eq:decmap} and gives
\begin{align}
  \avg{J_{x,y}}_\Gamma &= e^{-\Gamma t} \avg{J_{x,y}}_0 \;,\\
  (\Delta J_{x,y})^2_\Gamma &= e^{-2\Gamma t}(\Delta J_{x,y})^2_0 +
  \frac{J}{2}(1-e^{-2\Gamma t})\;.
\end{align}
Here a subscript $\Gamma$ denotes the value with dephasing, and a
subscript $0$ without.  It is now easy to see that under this model
of decoherence, for either of the measurements that we are
considering, the sensitivity takes the form
\begin{align}
\label{eq:vd}
\delta \gamma_\Gamma^2&=
\delta\gamma^2 +
\frac{J(e^{2\Gamma t}-1)}{2\nu(d\avg{J_{x,y}}_0/d\gamma)^2}\nonumber\\
&=\delta\gamma^2
\left(1+\frac{J(e^{2\Gamma t}-1)}{2(\Delta J_{x,y})^2_0}\right)\;.
\end{align}

To assess the effects of decoherence, we now focus on $J_y$
measurements, and we assume that through an adaptive feedback
procedure like that sketched in Sec.~\ref{sec:mea}, we are operating
well within the central fringe, i.e., $\gamma t$ is somewhat smaller
than $\pi/4J\cos\beta$.  Inserting the $\phi=0$ values from
Eqs.~\eqref{eq:DeltaJy} and~\eqref{eq:yapp} into Eq.~\eqref{eq:vd}
yields a sensitivity
\begin{equation}
\delta \gamma_\Gamma=\frac{e^{\Gamma t}}{t\,\sqrt\nu}
\frac{1}{\sqrt{2}J^{3/2}|\sin2\beta|}\;.
\end{equation}
If we now let $T=\nu t$ be the total time available for measurements
involving $\nu$ probes, the optimal value of $t$, found by maximizing
$e^{\Gamma t}/\sqrt t$ is $t=\tau_2/2$, gives a sensitivity
\begin{equation}
\delta\gamma_\Gamma=
\sqrt{\frac{e}{T\tau_2}}\frac{1}{ J^{3/2}|\sin2\beta|}\;.
\end{equation}
This result assumes that each probe can be processed in a time
$\tau_2/2$, but within this constraint, the scaling is the same
$\cO(J^{-3/2})$ scaling that applies in the absence of decoherence.
This is to be contrasted with entangled inputs, where uncorrelated
phase decoherence degrades the scaling from $\cO(J^{-2})$ to the
$\cO(J^{-3/2})$ characteristic of product inputs.

These arguments hold for general symmetric $k$-body Hamiltonians,
giving a sensitivity scaling $\cO(n^{-k+1/2})$ for initial product
states subjected to uncorrelated phase decoherence.  This is the same
scaling achieved by initial optimal entangled states under this
decoherence model~\cite{rey07a}.  On the other hand, the use of
product states with $k$-body Hamiltonians for $k\ge2$ can surpass
both the standard quantum limit and the Heisenberg limit, even in the
presence of phase decoherence.

\section{Conclusion \label{sec6}}

The possibility of using nonlinear Hamiltonians has the potential to
open up a new frontier in quantum metrology.  Quantum metrology has
traditionally focused on linear Hamiltonians of the form $\gamma
J_z=\gamma\sum_{j=1}^n Z_j/2$.  The main technical challenge has been
to improve on the standard quantum limit for determining the
parameter $\gamma$, which scales as $\cO(n^{-1/2})$ and can be
attained relatively easily using product input states and separable
measurements.  The goal of linear quantum metrology has been to
achieve the Heisenberg limit for determining $\gamma$, which scales
as $\cO(n^{-1})$ and requires the use of highly entangled input
states.  Nonlinear coupling Hamiltonians of the form $J_z^k$ offer
the possibility of further improvements in scaling.  With the same
highly entangled input states, nonlinear Hamiltonians can achieve a
scaling $\cO(n^{-k})$.  More importantly, they provide
$\cO(n^{-k+1/2})$ scalings, better than the Heisenberg limit, for
input product states and separable measurements.  We expect that the
generalized quantum metrology of nonlinear Hamiltonians will lead to
new experiments---and, ultimately, to new devices---that take
advantage of the enhanced scaling, which is available using the
experimentally accessible tools of product-state inputs and separable
measurements.

A notable feature of generalized quantum metrology is that the
enhanced scalings available with product-state inputs do not rely on
the entanglement produced by the nonlinear Hamiltonian.  We reach
this conclusion in this paper from a detailed analysis of the $k=2$
case, in the course of which we formulate an approximate
coherent-state model of the time evolution, which applies during the
period of enhanced sensitivity.  In the model, a coherent state that
makes an angle $\beta$ to the $z$ axis rotates with angular velocity
$2\gamma J\cos\beta$.  The increased rotation rate, larger by a
factor of $2J\cos\beta$ than for $k=1$, accounts for the enhanced
sensitivity.  Since coherent states are product states, this
indicates that entanglement plays no role in the enhanced
sensitivity, and it accounts for the robustness we find in the
presence of phase decoherence.

Although these conclusions emerge here from the $k=2$ analysis in
this paper, it is not hard to extend the coherent-state model to
arbitrary $k$.  Given the input state~\eqref{eq:is}, the state
at time $t=\phi/\gamma$ becomes
\begin{equation}\label{eq:tsk}
\ket{\Psi_\beta(t)} = e^{-i\phi J_z^k}\ket{\Psi_\beta}=
\sum_m d_m e^{-i\phi m^k}\ket{J,m}\;.
\end{equation}
The squares of the Wigner rotation-matrix elements $d_m$ of
Eq.~\eqref{eq:djm} are a binomial distribution, which for large $J$,
approaches a narrow Gaussian, centered at $m=\avg{J_z}=J\cos\beta$,
with half-width~$\sqrt{2J+1}\sin\beta$.  This encourages us to
approximate $m^k$ in the phases of Eq.~\eqref{eq:tsk} as
$(J\cos\beta+\Delta m)^k\simeq
(J\cos\beta)^k+k(J\cos\beta)^{k-1}\Delta m$, giving
\begin{align}
\ket{\Psi_\beta(t)}&=
e^{i\phi(k-1)(J\cos\beta)^k}
\sum_m d_m e^{-i\phi k(J\cos\beta)^{k-1}m}\ket{J,m}\nonumber\\
&=e^{i\phi(k-1)(J\cos\beta)^k}e^{-i\phi k(J\cos\beta)^{k-1}J_z}e^{-i\beta J_y}\ket{J,J}\;.
\label{eq:kcs}
\end{align}
This is an angular-momentum coherent state at angle $\beta$ to the
$z$ axis, rotating about the $z$ axis with angular velocity $\gamma
k(J\cos\beta)^{k-1}$, which is the same enhanced rotation rate that
we found in the very short-time analysis of Sec.~\ref{sec4}.  The
approximation leading to the coherent state~\eqref{eq:kcs} thus
extends to arbitrary~$k$ the uniform-fringe approximation, formulated
for $k=2$ in Sec.~\ref{sec5}.  The fringes have
width~$\pi/k(J\cos\beta)^{k-1}$, and the approximation provides a
reasonable description of the first and second moments of $J_x$ and
$J_y$ so long as $(J\cos\beta)^{k-2}\phi\sqrt J|\sin\beta|\ll1$.

The enhanced rotation rate is responsible for the improved scaling,
and just as for $k=2$, the coherent-state model indicates that the
entanglement generated by the nonlinear Hamiltonian plays no role in
the enhancement.  In separate work, to be published elsewhere, we
extend these ideas.  We investigate in more detail the entanglement
generated by the nonlinear Hamiltonian, quantifying it using standard
entanglement measures and showing that the enhanced sensitivity with
initial product states can be achieved with a vanishing amount of
entanglement.

\vspace{-.2cm}
\acknowledgements

This work was supported in part by Office of Naval Research Grant
No.~N00014-07-1-0304 and by the National Nuclear Security
Administration of the U.S.~Department of Energy at Los Alamos National
Laboratory under Contract No.~DE-AC52-06NA25396. EB acknowledges
financial support from the the Catalan government, contract CIRIT SGR-
00185, and from the Spanish MEC through contracts FIS2005-01369, QOIT
(Consolider-Ingenio 2010) and travel grant PR2007-0204. He thanks the
Department of Physics and Astronomy of the University of New Mexico
for hospitality.

\begin{appendix}

\section{Symmetric Hamiltonian without self-interaction terms}
\label{appA}

The symmetric $k$-body coupling without self-interactions is
described by the Hamiltonian
\begin{align}
  \widetilde{H} &= \sum^n_{(a_1,\ldots,a_k)} h_{a_1}\cdots h_{a_k}\nonumber\\
  &= k!\!\!\sum^n_{a_1<a_2<\ldots<a_k} h_{a_1}\cdots h_{a_k}\nonumber\\
  &=\sum^n_{a_1,\ldots,a_k} h_{a_1}\cdots h_{a_k}\nonumber\\
  &\phantom{=x}-\binom{k}{2}\!\!\sum_{(a_1,\ldots,a_{k-1})}h_{a_1}\cdots h_{a_{k-2}}h_{a_{k-1}}^2
  +\cdots\;,
  \label{eq:tildeH}
\end{align}
where in the second form we use the expansions of Sec.~\ref{S:exp} to
relate $\widetilde H$ to the Hamiltonian with self-interactions, plus
corrections of which we give only the first.   The analysis of the
QCRB~\eqref{eq:semi-norm} relies on finding the largest and smallest
eigenvalues of $\widetilde{H}$.  Since we are only interested in the
leading-order scaling of the QCRB, Eq.~\eqref{eq:tildeH} shows that
the analysis proceeds exactly as in the corresponding analysis for
$H$ in Sec.~\ref{sec2}, with the proviso that the results are only
good to leading order in $n$.  Thus for $\widetilde H$, the
QCRB~\eqref{eq:semi-norm}, which can be achieved by using an
appropriate initial entangled state, scales as $\cO(n^{-k})$.

For the case of initial product states for the probe, we proceed as
in the comparable analysis of $H$ in Sec.~\ref{S:exp}.  Using the
same trick of artfully switching between restricted and unrestricted
sums, we can write
\begin{align}
\langle\widetilde H\rangle&=
\sum_{(a_1,\ldots,a_k)}\avg{h_{a_1}}\cdots\avg{h_{a_k}}\nonumber\\
&=\sum_{a_1,\ldots,a_k}\avg{h_{a_1}}\cdots\avg{h_{a_k}}\nonumber\\
&\phantom{=x}
-\binom{k}{2}\!\!\sum_{(a_1,\ldots,a_{k-1})}
\avg{h_{a_1}}\cdots\avg{h_{a_{k-2}}}\avg{h_{a_{k-1}}}^2\nonumber\\
&\phantom{=x}+\cO(n^{k-2})\;,
\end{align}
\begin{align}
\langle\widetilde H^2\rangle&=
\sum_{(a_1,\ldots,a_{2k})}\avg{h_{a_1}}\cdots\avg{h_{a_{2k}}}\nonumber\\
&\phantom{=x}
+k^2\!\!\sum_{(a_1,\ldots,a_{2k-1})}
\avg{h_{a_1}}\cdots\avg{h_{a_{2k-2}}}\avg{h_{a_{2k-1}}^2}\nonumber\\
&\phantom{=x}+\cO(n^{k-2})\;,
\end{align}
\begin{align}
\langle\widetilde H\rangle^2&=
\sum_{(a_1,\ldots,a_{2k})}\avg{h_{a_1}}\cdots\avg{h_{a_{2k}}}\nonumber\\
&\phantom{=x}
+k^2\!\!\sum_{(a_1,\ldots,a_{2k-1})}
\avg{h_{a_1}}\cdots\avg{h_{a_{2k-2}}}\avg{h_{a_{2k-1}}}^2\nonumber\\
&\phantom{=x}+\cO(n^{k-2})\;.
\end{align}
The resulting variance of $\widetilde H$,
\begin{align}
  (\Delta\widetilde H)^2 &=
  k^2\!\!\!\sum_{(a_1,\ldots,a_{2k-1})}
  \avg{h_{a_1}}\cdots\avg{h_{a_{2k-2}}}\Delta h_{a_{k-1}}^2\nonumber\\
  &\phantom{=x}+\cO(n^{2k-2})
  \;,
\end{align}
is the same, to leading order in $n$, as the variance~\eqref{eq:varHS}
of $H$, so the remaining analysis of the optimal scaling and initial
product states proceeds exactly as in Sec.~\ref{sec3}.

\section{First and second moments of $\vec J$}
\label{appB}

In this Appendix we derive the first and second moments of the
components of the angular momentum $\vec J$ for the Hamiltonian
$H_\gamma=\gamma J_z^2$ considered in Sec.~\ref{sec5}.

Since $J_z$ commutes with the Hamiltonian, it is a constant of the
motion, and its moments are at all times those of the initial
coherent state:
\begin{align}
\avg{J_z}_\phi&=J\cos\beta\;,\\
\avg{J_z^2}_\phi&=J^2\cos^2\!\beta+\frac{J}{2}\sin^2\!\beta\;.
\end{align}

In finding the first and second moments involving the equatorial
components of $\vec J$, it is convenient to work in terms of the
angular-momentum raising and lowering operators, $J_\pm=J_x\pm iJ_y$,
which act according to $J_\pm \ket{J,m} = \Gamma_m^\pm \ket{J,m \pm
1}$, where $\Gamma_m^{\pm}=\sqrt{(J \mp m)(J \pm m+1)}$.

For the evolved state~\eqref{eq:ts}, we can write
\begin{align}
\avg{J_+}_\phi &=
\sum_{m,m'=-J}^J d_m d_{m'}e^{i\phi(m'^2-m^2)}\Gamma_m^+\delta_{m',m+1} \nonumber \\
&=\sum_{m=-J}^J d_m d_{m+1}\Gamma_m^+e^{i\phi(2m+1)})\nonumber\\
&=\cot(\beta/2)\sum_{m=-J}^J(J-m)d_m^2 e^{i\phi(2m+1)}\;,
\end{align}
where the last line uses
\begin{equation}
\label{dprop}
d_{m+1}=\sqrt{\frac{J-m}{J+m+1}}\cot(\beta/2)d_m\;.
\end{equation}
Using Eq.~\eqref{eq:djm} and a derivative of the binomial formula,
\begin{equation}
\label{eq:jm}
\sum_{m=-J}^J (J-m)
\binom{2J}{J-m}a^{J+m}b^{J-m} = 2Jb(a+b)^{2J-1}\;,
\end{equation}
we obtain Eq.~\eqref{eq:Jplus}:
\begin{align}
\label{eq:Jplusa}
\avg{J_+}_\phi &=
\cot(\beta/2)e^{i\phi}
\sum_{m=-J}^J(J-m)\binom{2J}{J-m} \nonumber \\
&\phantom{\cot}\times\big[e^{i\phi}\cos^2(\beta/2)\big]^{J+m}
\big[e^{-i\phi}\sin^2(\beta/2)\big]^{J-m}\nonumber\\
&=J\sin\beta(\cos\phi+i\sin\phi\cos\beta)^{2J-1}\;.
\end{align}

The evaluation of the remaining second moments proceeds along the
same lines.  The correlation between $J_z$ and $J_x$ or $J_y$ is
conveniently expressed by
\begin{align}
\frac{1}{2}&\avg{J_z J_++ J_+ J_z}_\phi\nonumber\\
&=\sum_{m=-J}^J d_m d_{m+1}(m+1/2)\Gamma_m^+e^{i\phi(2m+1)})\nonumber\\
&=(1/2-J)\avg{J_+}_\phi\nonumber\\
&\phantom{=x}+\cot(\beta/2)\sum_{m=-J}^J(J^2-m^2)d_m^2 e^{i\phi(2m+1)}\nonumber\\
&=\frac{J(2J-1)}{2}\sin\beta(\cos\phi\cos\beta+i\sin\phi)\nonumber\\
&\phantom{=x}\times(\cos\phi+i\sin\phi\cos\beta)^{2(J-1)}\;,
\end{align}
where we use
\begin{align}
\label{eq:j2m2}
\sum_{m=-J}^J&(J^2-m^2)\binom{2J}{J-m}a^{J+m}b^{J-m}\nonumber\\
&=2J(2J-1)ab(a+b)^{2(J-1)}\;,
\end{align}
To find the second moments that involve only the equatorial
components, we use
\begin{subequations}
\begin{align}
&J_x^2+J_y^2=\frac{1}{2}(J_+ J_- + J_- J_+)\;,\\
&J_x^2-J_y^2=\frac{1}{2}(J_+^2 + J_-^2)\;,\\
&J_xJ_y+J_yJ_x=\frac{1}{2i}(J_+^2 - J_-^2)\;,
\end{align}
\end{subequations}
from which we get
\begin{subequations}
\label{eq:Jxysqa}
\begin{align}
&\avg{J_{x,y}^2}_\phi=
\frac{1}{4}\avg{J_+ J_- + J_- J_+}_\phi
\pm\frac{1}{2}\mathrm{Re}\big(\avg{J_+^2}_\phi\big)\;,\\
&\avg{J_xJ_y+J_yJ_x}_\phi=\mathrm{Im}\big(\avg{J_+^2}_\phi\big)\;.
\label{eq:JxJya}
\end{align}
\end{subequations}
Thus we calculate Eq.~\eqref{eq:JplusJminus},
\begin{align}
\frac{1}{2}\avg{J_+ J_- + J_- J_+}_\phi&=
\sum_{m=-J}^J(J+J^2-m^2)d_m^2\nonumber\\
&=J+\frac{J(2J-1)}{2}\sin^2\!\beta\;,
\end{align}
where we use Eq.~\eqref{eq:j2m2}, and we calculate
Eq.~\eqref{eq:Jplussq},
\begin{align}
\label{eq:Jplussqa}
\avg{J_+^2}_\phi&=
\sum_{m=-J}^Jd_m d_{m+2}e^{4i\phi(m+1)}\Gamma_{m+1}^+\Gamma_m^+\nonumber\\
&=\cot^2(\beta/2)\sum_{m=-J}^J(J-m)(J-m-1)d_m^2 e^{4i\phi(m+1)}\nonumber\\
&=\frac{J(2J-1)}{2}\sin^2\!\beta(\cos2\phi+i\sin2\phi\cos\beta)^{2(J-1)}\;,
\end{align}
where we use
\begin{align}
\label{eq:jmjm1}
\sum_{m=-J}^J&(J-m)(J-m-1)\binom{2J}{J-m}a^{J+m}b^{J-m}\nonumber\\
&=2J(2J-1)b^2(a+b)^{2(J-1)}\;.
\end{align}
The equatorial second moments listed in Eq.~\eqref{eq:Jxysq2} and the
cross moment,
\begin{align}
\frac{1}{2}&\avg{J_xJ_y+J_yJ_x}_\phi=
\frac{1}{2}\mathrm{Im}\big(\avg{J_+^2}_\phi\big)\nonumber\\
&=\frac{J(2J-1)}{4}\sin^2\!\beta\,R^{2(J-1)}\sin[2(J-1)\Theta]\;,
\end{align}
follow from inserting these results into Eqs.~\eqref{eq:Jxysqa}.

We now make the uniform-fringe approximation of Sec.~\ref{sec5},
keeping only the fringe terms near $\phi=0$.  This approximation
requires that $\sqrt J\phi|\sin\beta|\ll1$, but allows
$J\phi\cos\beta$ to be considerably bigger than 1 when
$J\cot^2\!\beta$ is large.  The resulting first and second moments,
\begin{subequations}
\begin{align}
&\avg{J_z}_\phi=J\cos\beta\;,\\
&\avg{J_+}_\phi\simeq J\sin\beta\,e^{2iJ\phi\cos\beta}\;,\\
&\avg{J_z^2}_\phi=J^2\cos^2\!\beta+\frac{J}{2}\sin^2\!\beta
=\frac{J}{2}+\frac{J(2J-1)}{2}\cos^2\!\beta\;,\\
&\frac{1}{2}\avg{J_z J_++ J_+ J_z}_\phi\simeq
\frac{J(2J-1)}{2}\sin\beta\cos\beta\,e^{2iJ\phi\cos\beta}\;,\\
&\avg{J_x^2}_\phi\simeq
\frac{J}{2}
+\frac{J(2J-1)}{2}\sin^2\!\beta\cos^2(2J\phi\cos\beta)\;,\\
&\avg{J_y^2}_\phi\simeq
\frac{J}{2}
+\frac{J(2J-1)}{2}\sin^2\!\beta\sin^2(2J\phi\cos\beta)\;,\\
&\frac{1}{2}\avg{J_xJ_y+J_yJ_x}_\phi\nonumber\\
&\phantom{\frac{1}{2}x}
\simeq\frac{J(2J-1)}{2}\sin^2\!\beta\sin(2J\phi\cos\beta)\cos(2J\phi\cos\beta)\;,
\end{align}
\end{subequations}
have the unique form of an angular-momentum coherent state.  They
show that in the uniform-fringe approximation, the state is an
angular-momentum coherent state oriented at angle $\beta$ to the $z$
axis and rotating about the $z$ axis with angular velocity $2\gamma
J\cos\beta=2\gamma \avg{J_z}$.

\end{appendix}

\end{document}